\documentstyle[aps,twocolumn]{revtex}

\begin{document}
\draft
\title{Neutrino emission via the plasma process in a magnetized plasma}
\author{M.P. Kennett and D.B. Melrose}
\address{Special Research Centre for Theoretical Astrophysics, School of Physics, University of Sydney, NSW 2006, AUSTRALIA}
\date{\today}
\maketitle
\begin{abstract} 
Neutrino emission via the plasma process using the vertex formalism for QED in a strongly magnetized plasma is considered.  A new vertex function is introduced to include the axial vector part of the weak interaction.  Our results are compared with previous calculations, and the effect of the axial vector coupling on neutrino emission is discussed.  The contribution from the axial vector coupling can be of the same order as or greater than the vector vector coupling under certain plasma conditions.
\end{abstract}

\section{Introduction}
The decay of electromagnetic oscillations in a plasma into neutrinos is of interest as a stellar energy loss mechanism \cite{ggr,arw}.  The presence of a plasma allows the refractive index to be less than unity, which is necessary for the decay of a photon into a $\nu\bar{\nu}$ pair.  The resulting ``plasma process'' for neutrino emission, has been studied in the presence of an unmagnetized plasma \cite{arw,zaidi,dOnp,bs}.  There were also calculations in which the background medium was taken to be the magnetized vacuum; however the refractive index for the magnetized vacuum is always greater than unity, and it was assumed that the presence of some low density plasma could lead to an appropriate opening of phase space to allow the process to proceed \cite{G&N,vvs,Deraad,ioannisian}.  Here we perform a consistent calculation, in which the background plasma is included explicitly and thus the kinematic condition is not an {\it ad hoc} addition. 

The decay depends on the properties of the waves, and a magnetized plasma can support a variety of natural wave modes.  Canuto, Chiuderi and Chou \cite{can1,can2} considered the plasma process in a magnetized plasma and they considered several possible wave modes.  However their analysis neglected the axial vector aspect of the weak interaction and they also did not use the exact electron wavefunctions in a magnetic field.  These deficiencies raise doubts about the validity of their results at high magnetic field strengths.

In this paper we calculate the amplitude for the decay $\gamma \rightarrow \nu\bar{\nu}$ in a magnetized gas of electrons to $O(G_F)$.  We avoid the weaknesses in \cite{G&N,vvs,Deraad,can1,can2}, by using the exact electron wavefunctions in a magnetic field and including the effects of the background plasma.  In Sec.II, the formalism required to treat V-A interactions in a strongly magnetized plasma  is summarized and extended.  The formalism used is the vertex formalism \cite{MPI,MII,MPIII}, which allows both a momentum space representation for QED in a strong magnetic field, and a means to calculate the response tensors of a magnetized plasma.  In Sec.III the transition rate for the decay of a given wave mode is calculated.  Wave modes for a plasma with a cold electron distribution and for a thermal electron distribution are considered.  It is shown that the results of \cite{can1,can2} may be recovered with suitable approximations.  The implications of the axial vector aspect of the weak interaction are discussed, and the neutrino emission rates from different plasma modes are compared.  Natural units  with $\hbar = c = 1$ and Boltzmann's constant, $\kappa = 1$, are used throughout, and only standard neutrinos are considered.

We find that the presence of a magnetic field has very little effect on neutrino emission relative to that in an unmagnetized plasma except for magnetic fields close to the ``critical'' magnetic field strength $B_c = m^2/e = 4.41\times 10^9 \, {\rm T}$.  At high magnetic field strengths there is an enhancement of neutrino emission due to a large proportion of electrons being present in their lowest Landau orbital.  We derive a criterion for determining when the axial vector contribution is likely to be important for neutrino emission.

\section{V - A Interactions in a Magnetized Plasma}
\subsection{Vertex Formalism}
A systematic development of QED in a strong magnetic field was presented by Melrose and Parle \cite{MPI,MII,MPIII}.  A summary of the electron wavefunctions and vertex functions is contained in Appendix A.  The electron energy levels for a static background magnetic field of magnitude $B$ parallel to the 3-axis are $${\mathcal E}_q = (m^2 + p_{\parallel}^2 + 2neB)^{\frac{1}{2}},$$ where $\{q\}$ labels the set of quantum numbers which includes the parallel momentum, $p_{\parallel}$ and the Landau levels, $n=0,1,2,\ldots$ .  The ground state, $n=0,$ is a singlet state and the states $n>0$ are doubly degenerate due to two spin states.  In the Landau Gauge, the vector potential is ${\mathbf A}({\mathbf x}) = (0,Bx,0).$  The electron wave functions $\psi^{\epsilon}_q({\mathbf x})$ are eigenfunctions of a spin operator and the magnetic moment operator is chosen, cf. \cite{sok}:

\begin{equation}
\mbox{\boldmath$\hat{\mu}$}\equiv m{\mbox{\boldmath$\sigma$}} - i\gamma^1 {\mbox{\boldmath$\sigma$}} {\mathbf \times} ({\mathbf p} + e {\mathbf A}({\mathbf x})),
\end{equation}
where \boldmath$\sigma$\unboldmath\ denotes the Pauli spin matrices.  This spin operator commutes with both the Hamiltonian and radiative corrections to the Hamiltonian, and its eigenfunctions have symmetry between electron and positron states.  Here electron and positron states are labelled by $\epsilon$, electrons correspond to $\epsilon = 1$ and positrons to $\epsilon = -1$.
 
To allow a momentum representation, we use a vertex function $[\gamma_{q^{\prime}q} ^{\epsilon^{\prime}\epsilon}({\mathbf k})]^{\mu}$, defined in the following way \cite{MPI}:

\begin{eqnarray} \label{vertdef}
[\gamma _{q^{\prime}q}^{\epsilon^{\prime}\epsilon}({\mathbf k})]^{\mu} \equiv\frac{1}{V} \int d{\mathbf x} \exp(-i {\mathbf k}\cdot {\mathbf x}) \bar{\psi}^{\epsilon^{\prime}}_{q^{\prime}}({\mathbf x}) \gamma^{\mu} \psi^{\epsilon}_q({\mathbf x}),
\end{eqnarray} 

\noindent where $V$ is the normalization volume.  The incoming electron has quantum numbers $\{\epsilon, q\}$, the outgoing electron has quantum numbers $\{\epsilon^{\prime}, q^{\prime}\}$ and the outgoing photon has 4-momentum $k^{\mu} = (\omega,{\mathbf k}) = (\omega, k_{\perp}\cos\psi,k_{\perp}\sin\psi,k_{\parallel})$.  From the definition of the vertex function, the following symmetry property is clear:

\begin{equation}
\label{sym}
([\gamma^{\epsilon^{\prime}\epsilon}_{q^{\prime}q} ({\mathbf k})]^{\mu})^* =
[\gamma^{\epsilon \epsilon^{\prime}}_{q q^{\prime}} (-{\mathbf k})]^{\mu}.
\end{equation} 

\noindent Since we also wish to treat the axial vector component of the weak interaction, we define an axial vector (AV) vertex function similarly to Eq.~(\ref{vertdef}), replacing $\gamma^{\mu}$ with $\gamma^{\mu}\gamma^5$.  The AV vertex function is denoted by $[\gamma _{q^{\prime}q}^{\epsilon^{\prime} \epsilon}({\mathbf k})]^{\mu}_5$. 

An electron - photon vertex corresponds to the standard vertex function, and an electron - Z boson vertex corresponds to a combination of the standard vertex function and the AV vertex function.  The V-A theory of weak interactions ignores Z and W boson propagators and considers only the charged and neutral currents at a point interaction.  The neutral current component of the interaction may be expressed using the standard and AV vertex functions, and hence the charged current component of the interaction may also be expressed in this manner through the use of a Fierz transformation (see e.g.\cite{nieves}).

Another vertex function, 
$[\Gamma _{q^{\prime}q}^{\epsilon^{\prime}\epsilon}({\mathbf k})]^{\mu}$, which is a gauge invariant part of $[\gamma _{q^{\prime}q}^{\epsilon^{\prime}\epsilon}({\mathbf k})]^{\mu}$, is identified due to the desirability of having a gauge invariant theory.  The gauge invariant part of the AV vertex function is identified as $[\Gamma^{\epsilon^{\prime}\epsilon}_{q^{\prime}q}({\mathbf k})]^{\mu}_5$.  (An example of the separation of gauge dependent and gauge independent terms is given in Appendix A).

Using the vertex formalism, we obtain a momentum space representation of the effective V-A interaction Lagrangian as:
\begin{eqnarray}
{\mathcal L}_{\rm ef{}f} & = & -\frac{G_F}{\sqrt{2}}
\bar{u}(q_1)\gamma_{\mu}(1 - \gamma^5)v(q_2) \nonumber \\ & & \times \ \left\{ {\mathcal A}[\Gamma^{\epsilon^{\prime}\epsilon}_{q^{\prime}q}({\mathbf k})]^{\mu} + {\mathcal B}[\Gamma^{\epsilon^{\prime}\epsilon}_{q^{\prime}q}({\mathbf k})]^{\mu}_5\right\},
\end{eqnarray}
where ${\epsilon,q}$ and ${\epsilon^{\prime},q^{\prime}}$ label the incoming and outgoing electron states, respectively, and $u(q_1)$ and $v(q_2)$ are the neutrino and antineutrino wavefunctions, respectively.  The Fermi constant is represented by $G_F$ and the constants ${\mathcal A}$ and ${\mathcal B}$ are given by ${\mathcal A} = 2\sin^2\theta_{\rm W} + \frac{1}{2}$ and ${\mathcal B} = -\frac{1}{2}$ for electron neutrinos, and by ${\mathcal A} = 2\sin^2\theta_{\rm W} - \frac{1}{2}$ and ${\mathcal B} = \frac{1}{2}$ for muon and tau neutrinos, where $\theta_{\rm W}$ is the Weinberg angle.  In the approximation where $\sin^2\theta_{\rm W} = \frac{1}{4}$, then ${\mathcal A} = 0$ for muon and tau neutrinos, so that only the axial vector component of the weak interaction contributes to their emission, as noted by \cite{ioannisian}.

The gauge invariant form of the AV vertex function is given in Appendix A.  When calculated using magnetic moment operator eigenfunctions, it obeys similar symmetry relations to the standard vertex function.  Explicitly these relations are

\begin{eqnarray}
([\Gamma^{\epsilon^{\prime}\epsilon}_{q^{\prime}q}({\mathbf k})]^{\mu}_5)^* & = & [\Gamma^{\epsilon\epsilon^{\prime}}_{qq^{\prime}}(-{\mathbf k})]^{\mu}_5,
\end{eqnarray}
\vspace{-0.81cm}
\begin{eqnarray}
\hspace*{0.75cm}[\Gamma^{-\epsilon^{\prime}-\epsilon}_{q^{\prime}q}(-{\mathbf k})]^{\mu}_5 & = &
(-1)^{l^{\prime}-l}[\Gamma^{\epsilon^{\prime}\epsilon}_{q^{\prime}q}({\mathbf k})]^{\mu}_5 .
\end{eqnarray}

\subsection{Response Tensors}

The standard linear response tensor for a plasma, $\Pi^{\mu\nu}(k)$ when written in covariant notation satisfies the equation

\begin{equation}
\label{cur}
j^{\mu}(k) = \Pi^{\mu}_{\;\;\nu}(k)A^{\nu}(k),
\end{equation}
where $j^{\mu}(k)$ is the induced 4-current and $A^{\mu}(k)$ is the fluctuating part of the electromagnetic field.  In the absence of a plasma, $\Pi^{\mu\nu}(k)$ is the vacuum polarization tensor.  Using the vertex formalism, one can introduce a medium using the electron occupation numbers, and then with the assumption that the occupation numbers of a state are independent of spin, the linear response tensor in a magnetic field becomes (\cite{MPIII}):

\begin{eqnarray}
\label{tens}
\Pi^{\mu\nu}(k) & = & -\frac{e^3 B}{2\pi}\sum_{n^{\prime}, n=0}^{\infty}\ \sum_{\epsilon^{\prime}, \epsilon=\pm}\nonumber \\ & & \times \  \int\frac{dp_{\parallel}}{2\pi} \frac{\{\frac{1}{2}(\epsilon^{\prime} - \epsilon) + \epsilon n^{\epsilon}_q -
\epsilon^{\prime}n^{\epsilon^{\prime}}_{q^{\prime}}\}}
{\omega - \epsilon{\mathcal E}_q + \epsilon^{\prime}{\mathcal E}_{q^{\prime}} + i0} T^{\mu\nu}_{\epsilon^{\prime}\epsilon}, 
\end{eqnarray}
where $T^{\mu\nu}_{\epsilon^{\prime}\epsilon}$ is the product of vertex functions summed over spin states, i.e.

\begin{eqnarray}
\label{summm}
T^{\mu\nu}_{\epsilon^{\prime}\epsilon} & = & \sum_{\sigma^{\prime}, \sigma =\pm} [\Gamma^{\epsilon^{\prime}\epsilon}_{q^{\prime}q}({\mathbf k})]^{\mu}
[\Gamma^{\epsilon^{\prime}\epsilon}_{q^{\prime}q}({\mathbf k})]^{\nu *}.
\end{eqnarray}

\noindent The result of the summation in Eq.~(\ref{summm}) is presented in Appendix A, the 3-tensor form of the response tensor was calculated in \cite{M74}, correcting the result of \cite{tsy62}, and the renormalized vacuum polarization tensor has been treated using the vertex formalism \cite{MS77}.  The infinitesimal imaginary term in the denominator of Eq.~(\ref{tens}) arises from the requirement that the response tensor be a causal function.  Note also that conservation of momentum is implicit through the relation $\epsilon^{\prime}p^{\prime}_{\parallel} = \epsilon p_{\parallel} - k_{\parallel}$.

The matrix element for the decay of a photon into a neutrino pair contains the product of a standard vertex function and an AV vertex function.  This allows one to identify an axial vector response function \cite{ioannisian}, which can be generalized in the same manner as the vacuum polarization tensor to include a medium.  Thus we have

\begin{eqnarray}
\Pi^{\mu\nu}_{\;\;5} (k) & = & -\frac{e^3 B}{2\pi}\sum_{n^{\prime} ,n=0}^{\infty}\ \sum_{\epsilon^{\prime}, \epsilon=\pm} \nonumber \\ & & \times \ \int\frac{dp_{\parallel}}{2\pi} \frac{\{\frac{1}{2}(\epsilon^{\prime} - \epsilon) + \epsilon n^{\epsilon}_q -
\epsilon^{\prime}n^{\epsilon^{\prime}}_{q^{\prime}}\}}
{\omega - \epsilon{\mathcal E}_q + \epsilon^{\prime}{\mathcal E}_{q^{\prime}} + i0} {}_5 T^{\mu\nu}_{\epsilon^{\prime}\epsilon},
\end{eqnarray}
with a similar sum over spin states

\begin{eqnarray}
{}_5 T^{\mu\nu}_{\epsilon^{\prime}\epsilon} =  \sum_{\sigma^{\prime}, \sigma = \pm}[\Gamma^{\epsilon^{\prime}\epsilon}_{q^{\prime}q}({\mathbf k})]^{\mu}
[\Gamma^{\epsilon^{\prime}\epsilon}_{q^{\prime}q}({\mathbf k})]^{\nu *}_5.
\end{eqnarray}

The axial vector response for a magnetized plasma has not been written down previously to the best of our knowledge, however there have been calculations of $\pi^{\mu\nu}_{\;\;5}$ using the proper time formalism for the magnetized vacuum \cite{Deraad,ioannisian}.  The result of the sum over spin states is presented in Appendix A.

\section{Matrix Element}
There are two contributing diagrams to the amplitude for the plasma process for neutrino emission to $O(G_F^2)$.  These diagrams, and the diagram for the process when regarded as a V-A interaction are shown in Fig.~1.  For simplicity, only electron neutrinos are considered here.  Using the V-A theory of the weak interaction, the matrix elements for the W and Z diagrams contributing to the decay may be expressed in the same form using a Fierz transformation.  The matrix element is 

\begin{eqnarray}
\label{matx}
M_{fi} &=& -\frac{G_F}{\sqrt{2}e} \bar{u}(q_1)\gamma^{\nu}(1 - \gamma^5) v(q_2) \nonumber \\ & & \times \ 
({\mathcal A} \Pi_{\mu\nu}(k) + {\mathcal B}\Pi^{\;\;5}_{\mu\nu}(k))A^{\mu}(k),
\end{eqnarray}

\noindent where $A^{\mu}$ is the fluctuating part of the electromagnetic field.  For a magnetized plasma, the only difference in the matrix element Eq.~(\ref{matx}) from that for an unmagnetized plasma is the form of the electron wavefunctions.

The decay rate for a plasma mode is taken to be the transition probability per unit volume of ${\mathbf x} - {\mathbf k}$ space per unit time for the decay of a quantum of a plasma mode into a neutrino antineutrino pair.  Provided that the refractive index of  the plasma is close to unity then the decay rate $\Gamma_M$ for a mode $M$ can be written in the well known form (e.g. \cite{dOnp,bs,can1}):
 
\begin{equation}
\label{gam}
\Gamma_M = \frac{G^2_F R_M(k)}{6\pi e^2 \varepsilon_0 |\omega_M(k)|} (k^{\mu}k^{\nu}-k^2 g^{\mu\nu})Q_{\mu\nu},
\end{equation}
where $R_M(k)$ is the ratio of the electric energy to total energy in the wave, and  
\begin{eqnarray}
\label{qmunu}
Q_{\mu\nu} &=& |({\mathcal A}\Pi_{\sigma\mu}(k) + {\mathcal B}\Pi^{\;\; 5}_{\sigma\mu}(k)) e_M^{\sigma}(k)|^*\nonumber \\ & &  \times \ |({\mathcal A}\Pi_{\tau\nu}(k) + {\mathcal B}\Pi^{\;\;5}_{\tau\nu}(k))e_M^{\tau}(k)|.
\end{eqnarray}

The power emitted per unit volume for a given wave mode may be obtained by multiplying the decay rate for a given mode by the energy, $\omega_M$, of a photon in that mode and the plasmon occupation number $f(\omega_M)$.  Integrating over momentum space leads to the power emitted per unit volume by the decay of wave quanta into neutrinos as

\begin{equation}
\label{pow}
Q_M = \int\! \frac{d^3 {\mathbf k}}{(2\pi)^3} \omega_M\Gamma_M f(\omega_M).
\end{equation}

To calculate the neutrino emission from given plasma conditions, one proceeds through the following steps.  First, the electron distribution function is required to determine the plasma response.  Secondly, the plasma response is used to find the natural wave modes of the plasma.  The combined effects on the wave properties of the vacuum polarization of the magnetized vacuum and of the plasma response was discussed by \cite{KC}; but the vacuum contribution is ignored here.  Thirdly, the polarization vector, dispersion relation and response tensor are used to calculate the decay rate, Eq.~(\ref{gam}), which is then integrated in Eq.~(\ref{pow}) to determine the power emitted in neutrinos.  
The most difficult step in obtaining analytic results is solving for the wave modes in a magnetized plasma -- there are relatively few cases in which the modes are simple enough to allow computational ease.  However, given the relatively straightforward procedure to calculate the power emitted in neutrinos, there is the opportunity to obtain numerical rates for a large range of plasma conditions.  

A further simplification for computational ease is to take the long wavelength limit of the expressions for $\Pi^{\mu\nu}$ and $\Pi^{\mu\nu}_{\;\;5}$.  These expressions are presented in the Appendix.  Rather than using the response 4-tensor to determine the wave modes, it is convenient to use the dielectric 3-tensor which is related to the response 3-tensor by  $$K^i_{\;\;j} = \delta^i_{\;\;j} +\frac{1}{\varepsilon_0 \omega^2}\Pi^i_{\;\;j}.$$  The form of the dielectric tensor is the same as that for a cold plasma \cite{stix}

\begin{eqnarray}
\label{diel}
K^{i}_{\;\;j} = \left(\begin{array}{ccc} S & -iD & 0 \\ iD & S & 0 \\ 0 & 0 & P \end{array} \right).
\end{eqnarray}

\subsection{Neutrino Emission from a ``cold'' plasma}

Starting from Eq.~(\ref{diel}) for the dielectric tensor for a cold plasma, the equation for the refractive index $n$ takes the form

\begin{equation}
\label{abc}
An^4 - Bn^2 + C = 0,
\end{equation}
with the coefficients 
\begin{eqnarray}
A &=& S\sin^2\theta + P\cos^2\theta , \nonumber\\
B &=& (S^2-D^2)\sin^2\theta + S P (1+\cos^2\theta) ,\nonumber \\
C &=& P(S^2-D^2).
\end{eqnarray}

\noindent A specific solution $n=n_M$ of Eq.~(\ref{abc}) defines the mode $M$.  The general expression for the polarization vector of a mode may be expressed as \cite{M86}:

\begin{equation}
{\mathbf e}_M = \frac{K_M\mbox{\boldmath $\kappa$} + T_M{\mathbf t} + i{\mathbf a}}{(K_M^2 + T_M^2 + 1)^{\frac{1}{2}}},
\end{equation}
where \boldmath$\kappa$\unboldmath, ${\mathbf t}$ and ${\mathbf a}$ are given by 
\begin{eqnarray} 
\mbox{\boldmath $\kappa$ \unboldmath} = (\sin\theta,0,\cos\theta),
{\mathbf t} = (\cos\theta,0,-\sin\theta), \nonumber \\ {\mathbf a} = (0,1,0), 
\end{eqnarray}
and the coefficents $K_M$ and $T_M$ are given by
$$ K_M = \frac{(P-n_M^2)D\sin\theta}{An_M^2 - PS}, \quad T_M = \frac{DP\cos\theta}{An_M^2-PS}.$$
 
To calculate the natural modes from the dielectric tensor, the wavevector is taken to be ${\mathbf k} = |{\mathbf k}|\mbox{\boldmath $\kappa$} = |{\mathbf k}|(\sin\theta, 0, \cos\theta)$, where $\theta$ is the angle between the wavevector and the magnetic field.  Note that the choice of gauge here is the temporal gauge, thus the polarization vector for a mode $M$ takes the form $e_M^{\mu} = (0,{\mathbf e}_M)$, where ${\mathbf e}_M$ is the polarization 3-vector.  The polarization vectors take simple forms for the cases $\theta = 0$ and $\theta = \pi/2$.  For $\theta = 0$, there can be two circularly polarized modes (or only one if the other is evanescent), the ordinary and extraordinary modes, or only one longitudinal mode, the others being evanescent.  For more general angles of propagation, i.e. $\theta \ne 0,\pi/2$ the modes have neither purely longitudinal or purely transverse polarization.  The dispersion relations and polarization vectors for the modes at $\theta = 0$ and $\theta = \pi/2$ are given in Table 1.

A cold plasma electron distribution is 

\begin{equation}
\label{dist}
f^S_n({\mathcal E}_n) = \frac{4\pi^2}{eB}\left[n^+ + n^- \right]\delta_{n0} \delta(p_{\parallel}),
\end{equation}
where the $n^{\epsilon}({\mathcal E}_n)$ correspond to the number densities of electrons and positrons.  Equation (\ref{dist}) is a distribution in which all the electrons are in their lowest Landau orbital. Having substituted Eq.~(\ref{dist}) into the expression for the dielectric tensor, if we then take the classical limit ($\omega \ll m, eB \ll m^2$) and assume a purely electron plasma, we obtain the dielectric tensor of magnetoionic theory, which is exactly that used in previous investigations of the plasma process in a magnetized plasma \cite{can1,can2}.

Using the notation $X = {\omega_p^2}/{\omega^2}, Y = \Omega_e/\omega$, where $\omega_p$ is the plasma frequency (defined by $\omega_p^2 = e^2 n_e /\varepsilon_0 m $ ) and $\Omega_e$ is the electron cyclotron frequency, the magnetoionic theory gives \cite{M86}:

\begin{eqnarray}
{\displaystyle K^1_{\;\;1} = K^2_{\;\;2} = 1- \frac{X}{1-Y^2}, \quad K^1_{\;\;2} = -K^2_{\;\;1} = \frac{-iXY}{1-Y^2},} \\ {\displaystyle K^3_{\;\;3} = 1 -X,} \quad {\displaystyle  K^i_{\;\;j} = 0} \hspace{0.5cm} {\rm otherwise.} \nonumber
\end{eqnarray}

\subsubsection{Emission at $\theta = 0$}

The longitudinal mode at $\theta = 0$ is independent of the magnetic field and the decay rate is the same as the known value for an unmagnetized plasma \cite{arw,zaidi,dOnp,bs}.  The power emitted in the transverse modes at $\theta = 0$ can be written in a particularly simple way.  Taking $\lambda = 1$ to label the ordinary mode and $\lambda = -1$ to label the extraordinary mode,  the power emitted per unit solid angle is

\begin{equation}
\label{transv}
Q_{\lambda} = \frac{G_F^2}{384\pi^5\alpha}\int_{\omega_{\rm min}}^{\omega_{\rm max}} d\omega\ \omega^8 n_{\lambda} (1 - n_{\lambda}^2)^3 f(\omega),
\end{equation}
where $n_{\lambda}$ is the refractive index for the mode $\lambda$.  The frequencies $\omega_{\rm min}$ and $\omega_{\rm max}$ correspond to the frequencies at which the refractive index is $0$ and $1$ respectively -- for a cold classical plasma $\omega_{\rm max} = \infty$.  The frequencies here satisfy $\omega \ll m$, so this value is used for the upper cutoff.  Integrating from $\omega_{\rm min}$ might appear to contradict the assumption made in deriving Eq.~(\ref{gam}), that the refractive index be close to unity.  However, for almost all of the frequency range the refractive index is close to unity and thus the results obtained here are not compromised by this --- it is a far less serious approximation than the assumption of a cold plasma.  For a plasma described by the magnetoionic theory, Eq.~(\ref{transv}) reproduces the results of \cite{can1}.

\subsubsection{Neutrino Emission from modes at $\theta = \pi/2$}
The behavior of the ordinary and extraordinary modes at $\theta = \pi/2$ is less simple than for $\theta = 0$.  The axial vector part of the weak interaction can only couple to modes which have a component of their polarization vector parallel to the magnetic field.
The ordinary mode has such a component but the extraordinary mode does not.  For comparison, the power emitted by the ordinary mode per unit volume per unit solid angle is presented both with and without the axial vector coupling.  With the axial vector coupling one has

\begin{equation}
\label{ord1}
Q_o = \frac{G_F^2}{1536\pi^5\alpha}\int_{\omega_{\rm min}}^{\omega_{\rm max}} d\omega\  \omega^8 n_o (1-n_o^2)^2 (4 - 3n_o^2) f(\omega),
\end{equation}

\noindent and without the axial vector coupling one has

\begin{equation}
\label{ord2}
Q_o = \frac{G_F^2}{384\pi^5\alpha}\int_{\omega_{\rm min}}^{\omega_{\rm max}} d\omega\  \omega^8 n_o (1-n_o^2)^3 f(\omega).
\end{equation}
The power emitted in the extraordinary mode is

\begin{eqnarray}
\label{xtr}
Q_x & = & {\displaystyle \frac{G_F^2}{384\pi^5\alpha}\int_{\omega_{\rm min}}^{\omega_{\rm max}}} d\omega\  \omega^8 n_x (1 - n_x^2) \vspace{0.3cm} \\
& & \times \ \left[(S-1)^2 + D^2 - {\displaystyle \frac{4D^2 S(S-1)}{D^2 + S^2}}\right] f(\omega).
\end{eqnarray}
As can be seen in Fig.~2, the inclusion of the axial vector coupling leads to results which differ significantly from those obtained when it is ignored.  The effects are most pronounced for lower electron number densities, and there is only about a 25\% increase in emission close to the peak when AV effects are included.

Using the dispersion relation Eq.~(\ref{abc}) the plasma resonances (which correspond to $n^2 = \infty$) may be found from the equation $A/C = 0.$  In magnetoionic theory, the solutions to this equation are
\begin{equation}
\omega_{\pm}^2(\theta) = \frac{1}{2}(\omega_p^2 + \Omega_e^2)\pm \frac{1}{2}\{ (\omega_p^2 + \Omega_e^2)^2 - 4\omega_p^2\Omega_e^2\cos^2\theta\}^{\frac{1}{2}}.
\end{equation}
Canuto {\it et al} \cite{can2} claimed that $\omega_-^2(0) = \Omega_e^2 $ is a mode which can lead to enhanced neutrino emission at high plasma densities ($\rho > 10^{11} \ {\rm g \, cm}^{-3}$).  However, the plasma resonance does not satisfy the kinematic condition that the refractive index be less than unity, which is required for the plasma process to proceed --- hence no energy can be lost through this mechanism.  We conclude that there is no such enhanced emission at exceptionally high plasma densities.

\subsection{Neutrino Emission from a Thermal plasma}

To obtain analytic expressions for the energy loss in neutrinos from a thermal magnetized plasma one assumes a thermal form for the electron distribution in the expression for the response tensor Eq.~(\ref{tens}).  We make either a non-relativistic or semi-relativistic expansion of the resonant denominator in the tensor to simplify the analysis. 

\subsubsection{Role of the axial vector coupling}
For an unmagnetized plasma, it has been shown numerically that the AV contribution to energy loss via the plasma process is of the order of 0.01\% for temperatures below $10^{11} \,{\rm K}$ \cite{KIM}.  However, in a magnetized plasma, it is possible that the AV coupling can have a more significant effect on neutrino emission.  Physically this may be seen as follows: the AV coupling cannot affect processes in a system which has reflection symmetry; it requires that there be some axial vector in the system to which it can couple.  Although there is no such axial vector in a classical magnetized plasma (apart from the magnetic field, which is not relevant here), in a quantum magnetized plasma, the electronic ground state (the lowest Landau orbital) corresponds to a specific spin state, unlike all excited states which have two degenerate spin states.  A plasma with a significant fraction of its electrons in their lowest Landau orbitals thus has an appropriate axial vector that allows coupling to occur.

Hence we expect the AV component of the weak interaction to be important when a significant fraction of the electrons are in their lowest Landau level.  Consider a Fermi distribution of electrons 

\begin{equation}
\label{fermi}
f({\mathcal E}_q) = \frac{g_n}{\exp[({\mathcal E}_q - \mu)/T] + 1},
\end{equation}
where $g_n$ is the degeneracy of the $n$th energy level, $\mu$ is the chemical potential and $T$ is the temperature.  Taking the limit that T becomes large, Eq.~(\ref{fermi}) becomes a Maxwell - Boltzmann distribution, $$f({\mathcal E}_q) = g_n\exp[-({\mathcal E}_q - \mu)/T],$$ which when normalized in the non-relativistic limit gives \cite{dog}:

\begin{equation}
\label{ferdi}
f({\mathcal E}_q) = g_n\frac{4\pi^2 n_e}{eB}\frac{\tanh(\lambda/2)}{(2\pi mT)^{\frac{1}{2}}}\exp\left(-\frac{p_{\parallel}^2}{2mT} - n\lambda\right),
\end{equation}
where $\lambda = eB/mT$.  The normalization that is used is 

\begin{equation}
\sum_{n=0}^{\infty}\int dp_{\parallel} g_n f({\mathcal E}_q) = \frac{4\pi^2n_e}{eB}
\end{equation}
The most probable value of $n$ should be when the energy associated with parallel motion is of the same order as that associated with perpendicular motion, corresponding to equipartition of energy.  This occurs when $p_{\parallel}^2/2mT = n\lambda$.  The condition is 

\begin{equation}
\label{equi}
\frac{p_{\parallel}^2}{2mT} = \frac{neB}{mT},
\end{equation}
If one replaces $p_{\parallel}^2$ by $\langle p_{\parallel}^2 \rangle = mT$ and $n$ by $\langle n \rangle$, Eq.~(\ref{equi}) gives

\begin{equation} 
\langle n \rangle \simeq \frac{\langle p_{\parallel}^2 \rangle}{2e\hbar B} = \frac{m\kappa T}{2e\hbar B} \simeq 0.06 \frac{T}{B},
\end{equation}
where $T$ is the temperature in kelvin and $B$ is the magnetic field in tesla.  For a young, highly magnetized white dwarf star, one might expect the surface values of magnetic field and temperature to be $B \sim 10^5 \, {\rm T}$ and $ T \le 10^6 \, {\rm K}$ respectively, which gives $\langle n \rangle \simeq 0.6$, so a sizeable proportion of the electrons are in their lowest Landau orbital.  Hot, strongly magnetized white dwarfs or their precursors, and neutron stars, are objects which are likely to have their plasma neutrino emission affected by the presence of a strong magnetic field.

If we take $\langle n \rangle \leq 1$ to characterize when most of the electrons are in their lowest Landau orbital, then a criterion for whether the AV part of the weak interaction is important for neutrino emission is

\begin{equation}
\label{crit}
\left(\frac{B}{\rm\, T}\right) \geq 0.06 \left(\frac{T}{\rm\, K}\right).
\end{equation} 

Temperature and magnetic field regimes that occur in the interior of neutron stars and white dwarfs are compared with the criterion Eq.~(\ref{crit}) in Fig.~3.  Whilst there is a larger range of $B$ and $T$ conditions available for neutron stars, the electrons are almost certainly degenerate, in which case, the refractive index is greater than unity, and the plasma process is forbidden.  The white dwarf and neutron star internal conditions are taken from \cite{ggr}.

\subsubsection{Neutrino Emission}
To calculate the response tensor, start with the distribution function Eq.~(\ref{ferdi}) and substitute into Eq.~(\ref{tens}).  In the absence of positrons, the response tensor takes the form 

\begin{eqnarray}
\label{god}
\Pi^{\mu\nu}(k)  & = & -{\displaystyle \frac{e^3B}{4\pi^2}\sum\limits_{n,n^{\prime}=0}^{\infty}}{\displaystyle \int} dp_{\parallel}  \Bigg\{ {\displaystyle \frac{f^+({\mathcal E}_q) - f^+({\mathcal E}_{q^{\prime}})}{\omega - {\mathcal E}_q + {\mathcal E}_{q^{\prime}} + i0}T^{\mu\nu}_{++}} \nonumber \\ & & + {\displaystyle \frac{f^+({\mathcal E}_q)}{\omega - {\mathcal E}_q - {\mathcal E}_{q^{\prime}} + i0} T^{\mu\nu}_{-+}} \nonumber \\ & &  -\ {\displaystyle \frac{f^+({\mathcal E}_{q^{\prime}})}{\omega + {\mathcal E}_q + {\mathcal E}_{q^{\prime}} + i0} T^{\mu\nu}_{+-}} \Bigg\}.
\end{eqnarray}
$\Pi^{\mu\nu}_{\;\;5}$ may be found by replacing $T^{\mu\nu}_{\epsilon^{\prime}\epsilon}$ by ${}_5T^{\mu\nu}_{\epsilon^{\prime}\epsilon}$.

In making the non-relativistic approximation, it is assumed that the thermal energy of the electrons is much less than their rest mass energy,  i.e. $T \ll m \sim 6 \times 10^9 \, {\rm K}$ \cite{KC}.  Provided that one considers a plasma with $\Omega_e \ll m$ and modes such that $\omega \le \Omega_e$, so that  ${\mathcal E}_q + {\mathcal E}_{q^{\prime}} \ge 2m \gg \omega,$ the second and third denominators do not vanish.  There are three transitions that can occur in the highly magnetised plasma, all of which must be taken into account when calculating the plasma response.  There are processes in which an electron remains in the same Landau orbital after emission, i.e. $n=0, n^{\prime} = 0$, and there is also cyclotron emission ($n = 1, n^{\prime} = 0$) and cyclotron absorption ($n=0, n^{\prime} = 1$).  The first denominator has a zero, called a resonance, corresponding to either cyclotron emission or cyclotron absorption.  The resonant term is sensitive to finite temperature effects, but the two non-resonant terms are not \cite{dog}.  Hence for the two non-resonant terms, one may set the distribution function to be

\begin{eqnarray}
f({\mathcal E}_q) = \frac{4\pi^2 n_e}{eB}\delta(p_{\parallel})\delta_{n0}, 
\end{eqnarray}
These ``non-resonant'' contributions to the response tensors are shown in Appendix B. 

In the resonant terms we use a Maxwellian distribution of electrons in their lowest Landau orbital:
\begin{equation} 
f({\mathcal E}_q) = \frac{1}{\sqrt{2\pi mT}}\frac{4\pi^2n_e}{eB}\delta_{n0} \exp\left(-\frac{p_{\parallel}^2}{2mT}\right).
\end{equation}

\noindent For cyclotron emission the resonant denominator is   
\begin{equation}
\omega - {\mathcal E}_q + {\mathcal E}_{q^{\prime}} \simeq (\omega + \Omega_e) - \frac{p_{\parallel}k_{\parallel}}{m},
\end{equation}
and for cyclotron absorption we have
\begin{equation}
\omega - {\mathcal E}_q + {\mathcal E}_{q^{\prime}} \simeq (\omega - \Omega_e) - \frac{p_{\parallel}k_{\parallel}}{m}.
\end{equation}

\noindent The resonant response tensor becomes

\begin{eqnarray} 
\Pi^{\mu\nu}_{\rm res}(k) & = & -{\displaystyle \frac{e^2n_e}{\sqrt{2\pi mT}}\int_{-\infty}^{\infty}\! dp_{\parallel}\ }  {\displaystyle \exp\left(-\frac{p_{\parallel}^2}{2mT}\right)} \nonumber \\& & \times \  \left[ {\displaystyle \frac{T^{\mu\nu}_{++}(n=0,n^{\prime}=1)}{(\omega + \Omega_e) - p_{\parallel}k_{\parallel}/m}} -  {\displaystyle \frac{T^{\mu\nu}_{++}(n=1,n^{\prime}=0)}{(\omega - \Omega_e) - p_{\parallel}k_{\parallel}/m}} \right], \\
\end{eqnarray}
with the expression for ${\Pi^{\mu\nu}_{\;\;5}}_{\rm res}$ obtained by replacing $T^{\mu\nu}_{++}$ by ${}_5T^{\mu\nu}_{++}$.  This leads naturally to expressing the components of the tensors in terms of the plasma dispersion function $\bar{\phi}(z)$, defined by \cite{M86}  

\begin{equation}
\label{fizbuz}
 \bar{\phi}(z) \equiv \frac{z}{\sqrt{\pi}}\int_{-\infty}^{\infty}\! dt\ \frac{e^{-t^2}}{z - t},
\end{equation}
where $z = (\omega \pm \Omega_e)\sqrt{m}/k_{\parallel}\sqrt{2T}$.
To simplify the expressions obtained in Appendix B further, one can make use of the asymptotic expansion of $\bar{\phi}(z)$ for large $z$ \cite{M86}

\begin{equation}
\bar{\phi}(z) \simeq 1 + \frac{1}{2z^2} + \frac{3}{4z^4} + \ldots -i\sqrt{\pi}z e^{-z^2},
\end{equation}
taking only the highest order term in $z$ and ignoring the imaginary part.  This leads to Eq.~(\ref{final}) for the response tensor.  For waves propagating parallel to the magnetic field, the tensor reduces to  the same form as for a cold plasma; the power emitted in the ordinary and extraordinary modes is given by Eq.~(\ref{transv}), and the longitudinal mode is identical to that for an unmagnetized plasma.  Waves which are not propagating parallel to the magnetic field, lead to expressions for power emission which are more complicated than Eqs.~(\ref{transv})-(\ref{ord2}) and may have significant contributions from the axial vector coupling.

\section{Conclusions}

The calculations in this paper address a number of issues relating to the neutrino plasma process.  Firstly, the plasma process for neutrino emission has not previously been calculated taking into account strong magnetic field effects, plasma effects and the axial vector part of the weak interaction.  Our work takes a consistent approach to the inclusion of a plasma and the kinematic conditions under which the neutrino plasma emission process may occur, as opposed to the inconsistencies in previous calculations \cite{G&N,vvs,Deraad,ioannisian}. Secondly, the formalism for looking at weak processes in a strongly magnetized plasma, has not previously been able to deal with diagrams containing electron loops.  The axial vector vertex function and axial vector response tensor described here provide mathematical tools which can be used for such calculations.  Thirdly we have produced some analytic approximations to the power emitted in neutrinos from a volume of plasma with a given magnetic field, electron number density and temperature.  The exact results for the response tensors mean that these can be used to calculate numerical results for magnetic fields greater than the critical magnetic field -- this regime has not been investigated here.  We have also shown that contrary to the case of an unmagnetised plasma, the axial vector coupling can have a role in affecting neutrino emission via the plasma process, and we have suggested a simple criterion with which to estimate whether such axial vector effects are likely to be important.

The neutrino plasma process is related to neutrino Cerenkov radiation by a crossing symmetry (see \cite{ioannisian}), so that the results obtained here for the response functions can be used to study the Cerenkov process in plasmas with a refractive index greater than unity. 

The magnetic field dependence of the plasma process parallels the results found for neutrino dispersion in a strong field \cite{pedigree}, in that the results are relatively insensitive to the magnetic field.  The plasma process is only sensitive to the magnetic field for $B$ close to $B_c$ -- there is a much stronger dependence on temperature and electron number density than the magnetic field.  Hence unless one considers strongly magnetized plasmas, most of the expressions derived for unmagnetized plasmas are adequate.

There are several highly magnetized astrophysical environments where the plasma process may be of importance.  These are in the cooling of giant stars with highly magnetized cores, in the early stages of the evolution of a hot magnetized white dwarf.  The process may also be of importance for neutrino emission from neutron stars.  The enhanced neutrino emission due to the axial vector coupling in regions of stronger than average magnetic field might contribute to an anisotropic neutrino luminosity which has been suggested as a possible mechanism for the large proper motions of many pulsars.

\section{Acknowledgements}
The authors thank Stephen Hardy and Jeanette Weise for helpful comments.

\onecolumn
\begin{appendix}
 \section{}
The electron wavefunctions determined using the magnetic moment operator have been determined by \cite{MPI}, we display them here for convenience
\begin{equation}
\psi^{\epsilon}_q({\mathbf x},t) = e^{-i\epsilon{\mathcal E}_q t}\psi^{\epsilon}_q({\mathbf x}),
\end{equation}
where
\begin{eqnarray} 
\label{wvfn}
 \psi^{\epsilon}_q({\mathbf x})  & = & {\displaystyle \frac{\exp(i \epsilon p_y y + i \epsilon p_{\parallel} z)}{\{4{\mathcal E}_q{\mathcal E}_q^{0}
({\mathcal E}_q +{\mathcal E}_q^0)({\mathcal E}_q^0 +m)\}^{\frac{1}{2}}}} \times  \\  \\ 
& & \left\{ \hspace{0.37cm} \delta_{\epsilon,1}  \left[  \delta_{\sigma,1}  \left(  \begin{array}{c} ({\mathcal E}_q +{\mathcal E}_q^0)({\mathcal E}_q^0 +m) \upsilon_{n-1}(\xi) \\ -ip_n p_{\parallel}\upsilon_n(\xi) \\ p_{\parallel}({\mathcal E}_q^0 +m)\upsilon_{n-1}(\xi) \\ ip_n({\mathcal E}_q+{\mathcal E}_q^0) \upsilon_n(\xi) \end{array} \right)   
+ \delta_{\sigma,-1}  \left(  \begin{array}{c} -ip_n p_{\parallel}\upsilon_{n-1}(\xi) \\ ({\mathcal E}_q +{\mathcal E}_q^0)({\mathcal E}_q^0 +m) \upsilon_n(\xi) \\ -ip_n ({\mathcal E}_q +{\mathcal E}_q^0) \upsilon_{n-1}(\xi) \\ -p_{\parallel} ({\mathcal E}_q^0 +m)\upsilon_n(\xi) \end{array}  \right) \right] \right.  \vspace{0.2cm} 
\\ & & \left. +\ \delta_{\epsilon,-1}  \left[  \delta_{\sigma,1} \left( \begin{array}{c} p_{\parallel}({\mathcal E}_q^0 +m)\upsilon_{n-1}(\xi) \\ -ip_n({\mathcal E}_q+{\mathcal E}_q^0)\upsilon_n(\xi) \\ ({\mathcal E}_q+{\mathcal E}_q^0)({\mathcal E}_q^0 +m) \upsilon_{n-1}(\xi) \\ ip_n p_{\parallel} \upsilon_n(\xi) \end{array} \right) 
+ \delta_{\sigma,-1}  \left(   \begin{array}{c} ip_n ({\mathcal E}_q+{\mathcal E}_q^0)\upsilon_{n-1}(\xi) \\  -p_{\parallel}({\mathcal E}_q^0 +m)\upsilon_n(\xi) \\ ip_n p_{\parallel}\upsilon_{n-1}(\xi) \\ ({\mathcal E}_q+{\mathcal E}_q^0)({\mathcal E}_q^0 +m) \upsilon_n(\xi) \end{array}  \right) \right] \right\} , 
\end{eqnarray}

\noindent where the quantities $p_n, {\mathcal E}_q^0$ and ${\mathcal E}_q$ are given by:
\begin{eqnarray}
p_n  =  (2neB)^{\frac{1}{2}}, \quad
{\mathcal E}_q^0  =  (m^2 + p_n^2)^{\frac{1}{2}}, \quad
{\mathcal E}_q    =  ({{\mathcal E}_q^0}^2 + p_{\parallel}^2)^{\frac{1}{2}}.
\end{eqnarray}
In Eq.~(\ref{wvfn}), $\sigma$ is the spin quantum number which takes the values $\pm 1$ for spin up and spin down respectively, and $\epsilon$ is the sign of the energy.  If $p_{\parallel}$ is the $z$ component of momentum for an electron ($\epsilon = 1$), then $p_{\parallel}$ represents minus the $z$ component of momentum for a positron ($\epsilon = -1$).  The functions $\upsilon_n(\xi)$ are normalised simple harmonic oscillator wavefunctions of the form $$ \upsilon_n(\xi) \equiv \frac{H_n(\xi)\exp(-\frac{1}{2}\xi^2)}{(\pi^{1/2}2^n n!)^{1/2}},$$ where $H_n(\xi)$ is the $n$th Hermite polynomial and $$ \xi \equiv (eB)^{1/2}(x + \epsilon p_y /eB).$$

The separation of the vertex function into gauge dependent and gauge independent terms was given in \cite{MPI} for several choices of electromagnetic gauge.  
We write down their result for the Landau gauge 

\begin{eqnarray}
[\gamma^{\epsilon^{\prime}\epsilon}_{q^{\prime}q}({\mathbf k})]^{\mu} = \{(2\pi)^2/V(eB)^{\frac{1}{2}}\} \exp[ik_x(\epsilon p_y + \epsilon^{\prime} p_y^{\prime})/2eB]\delta(\epsilon p_y - \epsilon^{\prime} p_y^{\prime} - k_y) \delta(\epsilon p_z - \epsilon^{\prime} p_z^{\prime} - k_z) [\Gamma^{\epsilon^{\prime}\epsilon}_{q^{\prime}q}({\mathbf k})]^{\mu}.
\end{eqnarray}
A similar separation may be made for the AV vertex function.  One may write out the gauge invariant vertex function for the magnetic moment operator eigenfunctions as:

\begin{eqnarray}
[\Gamma _{q^{\prime}q}^{\epsilon^{\prime} \epsilon}({\mathbf k})]^{\mu}
&=& C^*_{q^{\prime}}C_q [ \delta_{\sigma^{\prime}\sigma} \{ \alpha^{\epsilon^{\prime}\epsilon}_{q^{\prime}q}(J^l_{l^{\prime} -l} + \rho^{\prime}_{n^{\prime}} \rho_n J^{l+\sigma}_{l^{\prime} -l}), \nonumber \\
& & \hspace{2.0cm} \epsilon \beta^{\epsilon^{\prime}\epsilon}_{q^{\prime}q}(-\rho_n \exp(i\sigma \psi)J^{l+\sigma}_{l^{\prime} -l-\sigma} - \rho^{\prime}_{n^{\prime}}\exp(-i\sigma \psi)J^l_{l^{\prime}-l+\sigma}), \nonumber\\
& & \hspace{2.0cm} i\epsilon \sigma \beta^{\epsilon^{\prime}\epsilon}_{q^{\prime}q} (\rho_n\exp(i\sigma \psi)J^{l+\sigma}_{l^{\prime} -l-\sigma} - \rho^{\prime}_{n^{\prime}}\exp(-i\sigma \psi)J^l_{l^{\prime}-l+\sigma}), \nonumber\\
& & \hspace{2.0cm} \eta^{\epsilon^{\prime}\epsilon}_{q^{\prime}q}(J^l_{l^{\prime} -l} + \rho^{\prime}_{n^{\prime}}\rho_n J^{l+\sigma}_{l^{\prime} -l}) \} \vspace{0.2cm} \nonumber \\
& & \hspace{0.2cm}-\epsilon \sigma  \delta_{\sigma^{\prime} -\sigma} \{  a^{\epsilon^{\prime}\epsilon}_{q^{\prime}q}(-\rho_n\exp(i\sigma \psi)J^{l+\sigma}_{l^{\prime} -l-\sigma} + \rho^{\prime}_{n^{\prime}}\exp(i\sigma \psi)J^l_{l^{\prime} -l-\sigma}), \nonumber \\
& & \hspace{2.0cm} \epsilon b^{\epsilon^{\prime}\epsilon}_{q^{\prime}q}(J^l_{l^{\prime} -l} - \rho^{\prime}_{n^{\prime}}\rho_n\exp(2i\sigma \psi)J^{l+\sigma}_{l^{\prime} -l-2\sigma}), \nonumber \\ 
& & \hspace{2.0cm} i\epsilon \sigma b^{\epsilon^{\prime}\epsilon}_{q^{\prime}q}(J^l_{l^{\prime} -l} + \rho^{\prime}_{n^{\prime}}\rho_n\exp(2i\sigma \psi)J^{l+\sigma}_{l^{\prime}-l-2\sigma}),  \\ 
& & \hspace{2.0cm} d^{\epsilon^{\prime}\epsilon}_{q^{\prime}q}(-\rho_n\exp(i\sigma \psi)J^{l+\sigma}_{l^{\prime} -l-\sigma} + \rho^{\prime}_{n^{\prime}}\exp(i\sigma \psi)J^l_{l^{\prime} -l-\sigma})\} ] , \nonumber
\end{eqnarray}

\noindent where the argument of the $J$ functions is the same as in Eq.~(\ref{AVvert}), and with the coefficients:
\begin{eqnarray}
\alpha^{\epsilon^{\prime}\epsilon}_{q^{\prime}q} & = & \delta_{\epsilon^{\prime}\epsilon} (1+\rho_{\parallel}^{\prime}\rho_{\parallel}) + \sigma \delta_{\epsilon^{\prime} -\epsilon}(\rho^{\prime}_{\parallel} + \rho_{\parallel}), \nonumber
\\
\beta^{\epsilon^{\prime}\epsilon}_{q^{\prime}q} & = & \delta_{\epsilon^{\prime} \epsilon} (1-\rho_{\parallel}^{\prime}\rho_{\parallel}) + \sigma \delta_{\epsilon^{\prime}-\epsilon}(\rho^{\prime}_{\parallel} - \rho_{\parallel}), \nonumber
\\
\eta^{\epsilon^{\prime}\epsilon}_{q^{\prime}q} & = & \delta_{\epsilon^{\prime} \epsilon} (\rho_{\parallel}^{\prime} +\rho_{\parallel}) + \sigma \delta_{\epsilon^{\prime} -\epsilon}(1+\rho^{\prime}_{\parallel}\rho_{\parallel}), \nonumber
\\ 
a^{\epsilon^{\prime}\epsilon}_{q^{\prime}q} & = & \delta_{\epsilon^{\prime} \epsilon} (\rho_{\parallel}^{\prime}+\rho_{\parallel}) - \sigma \delta_{\epsilon^{\prime} -\epsilon}(1+\rho^{\prime}_{\parallel}\rho_{\parallel}), \nonumber
\\
b^{\epsilon^{\prime}\epsilon}_{q^{\prime}q} & = & \delta_{\epsilon^{\prime} \epsilon} (\rho_{\parallel}^{\prime} -\rho_{\parallel}) - \sigma \delta_{\epsilon^{\prime} -\epsilon}(1-\rho^{\prime}_{\parallel}\rho_{\parallel}), \nonumber
\\
d^{\epsilon^{\prime}\epsilon}_{q^{\prime}q} & = & \delta_{\epsilon^{\prime}\epsilon} (1+\rho_{\parallel}^{\prime}\rho_{\parallel}) - \sigma \delta_{\epsilon^{\prime}-\epsilon}(\rho^{\prime}_{\parallel} + \rho_{\parallel}), 
\end{eqnarray}

\noindent and the abbreviations:
\begin{eqnarray}
C_q & \equiv & \left( \frac{({\mathcal E}_q +{\mathcal E}_q^0)({\mathcal E}_q^0 +m)}{4{\mathcal E}_q{\mathcal E}_q^0} \right)^{\frac{1}{2}} 
\{i\exp(i \psi )\}^l, \nonumber \\
C_{q^{\prime}} & \equiv & \left( \frac{({\mathcal E}_{q^{\prime}}+{\mathcal E}_{q^{\prime}}^0)({\mathcal E}_{q^{\prime}}^0 +m)}{4{\mathcal E}_{q^{\prime}} {\mathcal E}_{q^{\prime}}^0} \right)^{\frac{1}{2}} 
\{i\exp(i \psi )\}^{l^{\prime}}, 
\end{eqnarray}

\begin{eqnarray}
\rho_{\parallel} & \equiv & p_{\parallel}/({\mathcal E}_q+{\mathcal E}_q^0), \nonumber
\\
\rho_{\parallel}^{\prime} & \equiv & p_{\parallel}^{\prime}/({\mathcal E}_{q^{\prime}} + 
{\mathcal E}_{q^{\prime}}^{0}), \nonumber
\\
\rho_n & \equiv & p_n/({\mathcal E}_q^0+m), \nonumber
\\
\rho_{n^{\prime}}^{\prime} & \equiv & p_{n^{\prime}}/({\mathcal E}_{q^{\prime}}^0 +m). 
\end{eqnarray}

\noindent As noted by \cite{MPI}, the gauge invariant vertex function satisfies symmetry properties similar to Eq.~(\ref{sym}), provided that the magnetic moment operator eigenfunctions are used.  These symmetry properties are:

\begin{eqnarray}
([\Gamma^{\epsilon^{\prime}\epsilon}_{q^{\prime}q} ({\mathbf k})]^{\mu})^*  = 
[\Gamma^{\epsilon \epsilon^{\prime}}_{q q^{\prime}} (-{\mathbf k})]^{\mu}, 
\end{eqnarray}
\vspace{-0.5cm}
\begin{eqnarray}
\label{symma}
\hspace*{0.63cm} [\Gamma^{-\epsilon^{\prime}-\epsilon}_{q^{\prime}q} (-{\mathbf k})]^{\mu}  =  (-)^{l^{\prime}-l}
[\Gamma^{\epsilon^{\prime}\epsilon}_{q^{\prime}q} ({\mathbf k})]^{\mu},
\end{eqnarray}

\noindent with the symmetry property Eq.~(\ref{symma}) holding for suitably chosen phase factors.

The gauge invariant form of the AV vertex function is

\begin{eqnarray}
\label{AVvert}
[\Gamma _{q^{\prime}q}^{\epsilon^{\prime} \epsilon}({\mathbf k})]^{\mu}_5
& = & C_{q^{\prime}}C^*_q [ \delta_{\sigma^{\prime}\sigma} \{ \phi^{\epsilon^{\prime}\epsilon}_{q^{\prime}q}(J^l_{l^{\prime} -l} - \rho^{\prime}_{n^{\prime}} \rho_n J^{l+\sigma}_{l^{\prime}-l}), \nonumber
\\ 
& & \hspace{2.0cm} \pi^{\epsilon^{\prime}\epsilon}_{q^{\prime}q}(-\rho_n \exp(i\sigma \psi)J^{l+\sigma}_{l^{\prime}-l-\sigma} + \rho^{\prime}_{n^{\prime}}\exp(-i\sigma \psi)J^l_{l^{\prime}-l+\sigma}), \nonumber \\
& & \hspace{2.0cm} i\epsilon \sigma \pi^{\epsilon^{\prime}\epsilon}_{q^{\prime}q} (\rho_n\exp(i\sigma\psi)J^{l+\sigma}_{l^{\prime}-l-\sigma} +  \rho^{\prime}_{n^{\prime}}\exp(-i\sigma\psi)J^l_{l^{\prime}-l+\sigma}), \nonumber \\
& & \hspace{2.0cm} \theta^{\epsilon^{\prime}\epsilon}_{q^{\prime}q}(J^l_{l^{\prime}-l} - \rho^{\prime}_{n^{\prime}}\rho_n J^{l+\sigma}_{l^{\prime}-l}) \}\vspace{0.2cm} \nonumber \\
& & \hspace{0.2cm} -\epsilon \sigma  \delta_{\sigma^{\prime} -\sigma} \{ f^{\epsilon^{\prime}\epsilon}_{q^{\prime}q}(-\rho_n\exp(i\sigma \psi)J^{l+\sigma}_{l^{\prime}-l-\sigma}  - \rho^{\prime}_{n^{\prime}}\exp(i\sigma \psi)J^l_{l^{\prime}-l-\sigma}), \nonumber  \\
& & \hspace{2.0cm} \epsilon g^{\epsilon^{\prime}\epsilon}_{q^{\prime}q}(J^l_{l^{\prime}-l} + \rho^{\prime}_{n^{\prime}}\rho_n\exp(2i\sigma\psi)J^{l+\sigma}_{l^{\prime} -l-2\sigma}), \nonumber \\
& & \hspace{2.0cm} i\epsilon \sigma g^{\epsilon^{\prime}\epsilon}_{q^{\prime}q}(J^l_{l^{\prime}-l} - \rho^{\prime}_{n^{\prime}}\rho_n\exp(2i\sigma\psi)J^{l+\sigma}_{l^{\prime} -l-2\sigma}), \nonumber \\
& & \hspace{2.0cm} h^{\epsilon^{\prime}\epsilon}_{q^{\prime}q}(-\rho_n\exp(i\sigma \psi)J^{l+\sigma}_{l^{\prime}-l-\sigma} -  \rho^{\prime}_{n^{\prime}}\exp(i\sigma\psi)J^l_{l^{\prime} -l-\sigma}) \} ],
\end{eqnarray}

\noindent where $l = n - \frac{1}{2}(\sigma + 1)$ and $\sigma = \pm 1$ is the spin eigenvalue (note that for the ground state spin singlet $n=0$, the spin eigenvalue is $\sigma = -1$).  The $J$ functions have argument $k_{\perp}^2/2eB$, and are related to the generalized Laguerre polynomials (e.g.\cite{G&R}) via the relation

\begin{equation}
J^n_{\nu}(x) \equiv \left( \frac{n!}{(n+{\nu})!}\right)^{\frac{1}{2}}e^{-x/2}x^{\nu/2}
L^{\nu}_n (x) = (-)^{\nu}J^{n+\nu}_{-\nu}(x) .
\end{equation}
The properties of these functions have been summarized previously \cite{MPI}.  The coefficients for the AV vertex function are 

\begin{eqnarray}
\phi^{\epsilon^{\prime}\epsilon}_{q^{\prime}q} & = & \sigma \delta_{\epsilon^{\prime}\epsilon} (\rho_{\parallel}^{\prime}+\rho_{\parallel}) + \delta_{\epsilon^{\prime} -\epsilon}(1+\rho^{\prime}_{\parallel}\rho_{\parallel}), \nonumber
\\
\pi^{\epsilon^{\prime}\epsilon}_{q^{\prime}q} & = & \sigma \delta_{\epsilon^{\prime}\epsilon} (\rho_{\parallel}^{\prime}-\rho_{\parallel}) + \delta_{\epsilon^{\prime}-\epsilon}(1-\rho^{\prime}_{\parallel}\rho_{\parallel}), \nonumber
\\
\theta^{\epsilon^{\prime}\epsilon}_{q^{\prime}q} & = & \sigma\delta_{\epsilon^{\prime}\epsilon} (1+\rho_{\parallel}^{\prime}\rho_{\parallel}) + \delta_{\epsilon^{\prime} -\epsilon}(\rho^{\prime}_{\parallel}+\rho_{\parallel}), \nonumber
\\
f^{\epsilon^{\prime}\epsilon}_{q^{\prime}q} & = & -\sigma \delta_{\epsilon^{\prime}\epsilon} (1+\rho_{\parallel}^{\prime}\rho_{\parallel}) + \delta_{\epsilon^{\prime} -\epsilon}(\rho^{\prime}_{\parallel}+\rho_{\parallel}), \nonumber
\\
g^{\epsilon^{\prime}\epsilon}_{q^{\prime}q} & = & -\sigma\delta_{\epsilon^{\prime}\epsilon} (1-\rho_{\parallel}^{\prime}\rho_{\parallel}) + \delta_{\epsilon^{\prime} -\epsilon}(\rho^{\prime}_{\parallel}-\rho_{\parallel}), \nonumber
\\
h^{\epsilon^{\prime}\epsilon}_{q^{\prime}q} & = & -\sigma\delta_{\epsilon^{\prime}\epsilon} (\rho_{\parallel}^{\prime}+\rho_{\parallel}) +\delta_{\epsilon^{\prime} -\epsilon}(1+\rho^{\prime}_{\parallel}\rho_{\parallel}). 
\end{eqnarray}

\subsection{Response Tensors}
The sum over spin states for the linear response tensor and the vector axial response are presented below:

\begin{eqnarray}
\label{spin}
T^{00}_{\pm\epsilon\epsilon} & = &\frac{1}{2}\left\{1\pm\frac{(m^2 \pm p_{\parallel}^{\prime}p_{\parallel})}{{\mathcal E}_{q^{\prime}}{\mathcal E}_q}\right\}
[(J^n_{\nu})^2 + (J^{n-1}_{\nu})^2] \pm \frac{p_n p_{n^{\prime}}}{{\mathcal E}_{q^{\prime}}{\mathcal E}_q}J^n_{\nu} J^{n-1}_{\nu} \nonumber
,\\
T^{11}_{\pm\epsilon\epsilon}& = &\frac{1}{2}\left\{1\mp\frac{(m^2 \pm p_{\parallel}^{\prime}p_{\parallel})}{{\mathcal E}_{q^{\prime}}{\mathcal E}_q}\right\}
[(J^n_{\nu-1})^2 + (J^{n-1}_{\nu+1})^2] \pm \frac{p_n p_{n^{\prime}}}{{\mathcal E}_{q^{\prime}}{\mathcal E}_q}J^n_{\nu-1} J^{n-1}_{\nu+1} \nonumber
,\\
T^{22}_{\pm\epsilon\epsilon}& = &\frac{1}{2}\left\{1\mp\frac{(m^2 \pm p_{\parallel}^{\prime}p_{\parallel})}{{\mathcal E}_{q^{\prime}}{\mathcal E}_q}\right\}
[(J^n_{\nu-1})^2 + (J^{n-1}_{\nu+1})^2] \mp \frac{p_n p_{n^{\prime}}}{{\mathcal E}_{q^{\prime}}{\mathcal E}_q}J^n_{\nu-1} J^{n-1}_{\nu+1} \nonumber
,\\
T^{33}_{\pm\epsilon\epsilon}& = &\frac{1}{2}\left\{1\mp\frac{(m^2 \mp p_{\parallel}^{\prime}p_{\parallel})}{{\mathcal E}_{q^{\prime}}{\mathcal E}_q}\right\}
[(J^n_{\nu})^2 + (J^{n-1}_{\nu})^2] \mp \frac{p_n p_{n^{\prime}}}{{\mathcal E}_{q^{\prime}}{\mathcal E}_q}J^n_{\nu} J^{n-1}_{\nu} \nonumber
,\\
T^{01}_{\pm\epsilon\epsilon} & = & -\frac{\epsilon}{2} \left\{ \frac{p_n}{{\mathcal E}_q}(J^{n-1}_v J^n_{\nu-1} + J^n_v J^{n-1}_{\nu+1}) \pm \frac{p_{n^{\prime}}}{{\mathcal E}_{q^{\prime}}}(J^n_{\nu} J^n_{\nu-1} + J^{n-1}_{\nu} J^{n-1}_{\nu+1}) \right\} \nonumber
,\\
T^{02}_{\pm\epsilon\epsilon} & = & -\frac{i\epsilon}{2} \left\{ \frac{p_n}{{\mathcal E}_q}(J^{n-1}_v J^n_{\nu-1} - J^n_{\nu} J^{n-1}_{\nu+1}) \pm\frac{p_{n^{\prime}}}{{\mathcal E}_{q^{\prime}}}(J^n_{\nu} J^n_{\nu-1} - J^{n-1}_{\nu} J^{n-1}_{\nu+1}) \right\} \nonumber
,\\
T^{03} & = & \frac{1}{2} \left\{ \frac{p_{\parallel}^{\prime}}{{\mathcal E}_{q^{\prime}}} + \frac{p_{\parallel}}{{\mathcal E}_q} \right\} [(J^{n-1}_{\nu})^2 + (J^n_{\nu})^2] 
,\\
T^{12}_{\pm\epsilon\epsilon} & = & -\frac{i}{2} \left\{ 1 \mp \frac{(m^2 \pm p_{\parallel}^{\prime}p_{\parallel})}{{\mathcal E}_{q^{\prime}}{\mathcal E}_q} \right\} [(J^{n-1}_{\nu+1})^2 - (J^n_{\nu-1})^2)] \nonumber
,\\
T^{13}_{\pm\epsilon\epsilon} & = & -\frac{\epsilon}{2} \left\{ \frac{p_n p_{\parallel}^{\prime}}{{\mathcal E}_q{\mathcal E}_{q^{\prime}}}[J^n_{\nu-1}J^{n-1}_{\nu} + J^{n-1}_{\nu+1}J^n_{\nu}] \pm \frac{p_{n^{\prime}}p_{\parallel}}{{\mathcal E}_q{\mathcal E}_{q^{\prime}}}[J^n_{\nu} J^n_{\nu-1} + J^{n-1}_{\nu} J^{n-1}_{\nu+1}] \right\} \nonumber
,\\
T^{23}_{\pm\epsilon\epsilon} & = & -\frac{i\epsilon}{2} \left\{ \frac{p_n p_{\parallel}^{\prime}}{{\mathcal E}_q{\mathcal E}_{q^{\prime}}}[J^n_{\nu-1}J^{n-1}_{\nu} - J^{n-1}_{\nu+1}J^n_{\nu}] \pm \frac{p_{n^{\prime}}p_{\parallel}}{{\mathcal E}_q{\mathcal E}_{q^{\prime}}}[J^n_{\nu} J^n_{\nu-1} - J^{n-1}_{\nu} J^{n-1}_{\nu+1}] \right\}. \nonumber 
\end{eqnarray}
where $\nu = n^{\prime} - n$.  The remaining components may be constructed from the Onsager relations, which embody the requirements of time reversibility, and for a plasma with a static background magnetic field may be written in the form

\begin{eqnarray}
\label{lars}
\begin{array}{c}\Pi^{00}(\omega,-{\mathbf k})|_{-{\mathbf B}_0} = \Pi^{00}(\omega,{\mathbf k})|_{{\mathbf B}_0}, \quad \Pi^{0i}(\omega,-{\mathbf k})|_{-{\mathbf B}_0} = -\Pi^{i0}(\omega,{\mathbf k})|_{{\mathbf B}_0},  \\
\Pi^{ij}(\omega,-{\mathbf k})|_{-{\mathbf B}_0} = \Pi^{ji}(\omega,{\mathbf k})|_{{\mathbf B}_0}. \end{array}
\end{eqnarray}

The components of the 3-tensor part of $\Pi^{\mu\nu}_{\;\;5}$ may be constructed using the Onsager relations from the components given below, and the sum over spin states gives: 

\begin{eqnarray}
{}_5T^{00}& = &{}_5T^{33} =  \frac{1}{2}\left\{ \frac{p_{\parallel}^{\prime}}{{\mathcal E}_{q^{\prime}}} + \frac{p_{\parallel}}{{\mathcal E}_q} \right\} [(J^{n-1}_{\nu})^2 - (J^n_{\nu})^2] \nonumber
,\\
{}_5T^{11} & = &{}_5T^{22} =  \frac{1}{2}\left\{ \frac{p_{\parallel}^{\prime}}{{\mathcal E}_{q^{\prime}}} - \frac{p_{\parallel}}{{\mathcal E}_q} \right\} [(J^n_{\nu-1})^2 - (J^{n-1}_{\nu+1})^2]\nonumber
,\\
{}_5T^{01}_{\pm\epsilon\epsilon} & = & \frac{\epsilon}{2} \left\{\frac{p_{\parallel}^{\prime}p_n}{{\mathcal E}_{q^{\prime}}{\mathcal E}_q} [J^n_{\nu} J^{n-1}_{\nu+1} - J^{n-1}_{\nu} J^n_{\nu-1}] \mp \frac{p_{\parallel}p_{n^{\prime}}}{{\mathcal E}_{q^{\prime}}{\mathcal E}_q}[J^{n-1}_{\nu} J^{n-1}_{\nu+1}  - J^n_{\nu} J^n_{\nu-1}]\right\} \nonumber
,\\
{}_5T^{02}_{\pm\epsilon\epsilon} & = & -\frac{i\epsilon}{2}\left\{\frac{p_{\parallel}^{\prime}p_n}{{\mathcal E}_{q^{\prime}}{\mathcal E}_q} [J^n_{\nu} J^{n-1}_{\nu+1} + J^{n-1}_{\nu} J^n_{\nu-1}] \mp \frac{p_{\parallel}p_{n^{\prime}}}{{\mathcal E}_{q^{\prime}}{\mathcal E}_q}[J^{n-1}_{\nu} J^{n-1}_{\nu+1} + J^n_{\nu} J^n_{\nu-1}]\right\} \nonumber
,\\  
{}_5T^{03}_{\pm\epsilon\epsilon} & = & 
\frac{1}{2} \left\{ 1 \pm \frac{(m^2 \pm p_{\parallel}^{\prime}p_{\parallel})}{{\mathcal E}_{q^{\prime}}{\mathcal E}_q}\right\} [(J^{n-1}_{\nu})^2 - (J^n_{\nu})^2] \nonumber
,\\
{}_5T^{12} & = & \frac{i}{2} \left\{\frac{p_{\parallel}^{\prime}}{{\mathcal E}_{q^{\prime}}} - \frac{p_{\parallel}}{{\mathcal E}_q}\right\} [(J^n_{\nu-1})^2 + (J^{n-1}_{\nu+1})^2] \nonumber
,\\
{}_5T^{13}_{\pm\epsilon\epsilon} & = & \frac{\epsilon}{2} \left\{ \frac{p_n}{{\mathcal E}_q}[J^{n-1}_{\nu+1}J^n_{\nu} - J^n_{\nu-1}J^{n-1}_{\nu}] \mp \frac{p_{n^{\prime}}}{{\mathcal E}_{q^{\prime}}}[J^{n-1}_{\nu} J^{n-1}_{\nu+1} - J^n_{\nu} J^n_{\nu-1}] \right\} \nonumber
,\\
{}_5T^{23}_{\pm\epsilon\epsilon} & = & \frac{i\epsilon}{2} \left\{ \frac{p_n}{{\mathcal E}_q}[J^{n-1}_{\nu+1}J^n_{\nu} + J^n_{\nu-1}J^{n-1}_{\nu}] \mp \frac{p_{n^{\prime}}}{{\mathcal E}_{q^{\prime}}}[J^{n-1}_{\nu} J^{n-1}_{\nu+1} + J^n_{\nu} J^n_{\nu-1}] \right\} \nonumber
,\\
{}_5T^{30}_{\pm\epsilon\epsilon} & = & {}_5T^{03}_{\mp\epsilon\epsilon}. 
\end{eqnarray}

Most of the components of the tensor $\Pi^{\mu\nu}_{\;\;5}$ satisfy the Onsager relations in the form of Eq.~(\ref{lars}), specifically, the 3-tensor components and $\Pi^{01}_{\;\;5}$ and $\Pi^{02}_{\;\;5}$.  However, $\Pi^{03}_{\;\;5}$ does not exhibit the symmetry in Eq.~(\ref{lars}).  Physically, the reason for this is that the axial vector response violates parity.  If the failure to satisfy the Onsager relations were due to the non-conservation of the axial vector current, one would expect that the components to be affected would be associated with $k^{\nu}$.  This is not the case, so we can ascribe the failure to satisfy Eq.~(\ref{lars}) to parity violation. The Onsager relations are derived from time reversibility, but to write them in the form Eq.~(\ref{lars}), one also appeals to parity and the reality condition for Fourier transforms.  The tensor $\Pi^{\mu\nu}_{\;\;5}$ was calculated for a magnetized vacuum by \cite{ioannisian}, and we note that when Eq.(22) of \cite{ioannisian} is written in the co-ordinate system used here, it fails to satisfy the Onsager relations only for the $\Pi^{03}_{\;\;5}$ and $\Pi^{30}_{\;\;5}$ components.
 
\subsection{Plasma response in the long wavelength limit}
When one takes the long wavelength limit of Eq.~(\ref{tens}), the linear response tensor becomes (cf. \cite{M74}):

\begin{eqnarray}
\Pi^{11} & = & \Pi^{22} = -\frac{e^3 B}{2\pi}\sum_{n=0}^{\infty}\int\frac{dp_{\parallel}}{2\pi} \Bigg\{ (f^S_{n+1} - f^S_n) \frac{{\mathcal E}_{n+1} - {\mathcal E}_n}{\omega^2-({\mathcal E}_{n+1}- {\mathcal E}_n)^2}\left( 1 - \frac{(m^2 + p_{\parallel}^2)}{{\mathcal E}_{n+1}{\mathcal E}_n} \right) \vspace{0.2cm} \nonumber \\ 
& & +\ (f^S_{n+1} + f^S_n)\frac{{\mathcal E}_{n+1} + {\mathcal E}_n}{\omega^2 - ({\mathcal E}_{n+1} + {\mathcal E}_n)^2}\left(1+\frac{(m^2 + p_{\parallel}^2}{{\mathcal E}_{n+1}{\mathcal E}_n}\right) \Bigg\}, \nonumber \\
\Pi^{12} & = & \frac{ie^3 B}{2\pi}\sum_{n=0}^{\infty}\int\frac{dp_{\parallel}}{2\pi}\Bigg\{ (f^D_{n+1}-f^D_n)\frac{\omega}{\omega^2 - ({\mathcal E}_{n+1} - {\mathcal E}_n)^2}\left(1 - \frac{(m^2+p_{\parallel}^2)}{{\mathcal E}_{n+1}{\mathcal E}_n}\right) \nonumber \vspace{0.2cm} \\
& & +\ (f^D_{n+1} - f^D_n)\frac{\omega}{\omega^2 - ({\mathcal E}_{n+1} + {\mathcal E}_n)^2}\left(1+\frac{(m^2 +p_{\parallel}^2)}{{\mathcal E}_{n+1}{\mathcal E}_n}\right) \Bigg\}, \nonumber \\
\Pi^{33} & = & -4\frac{e^3 B}{2\pi}\sum_{n=0}^{\infty}\int\frac{dp_{\parallel}}{2\pi}\left\{ f^S_{n+1} \frac{{\mathcal E}_{n+1}}{\omega^2 -4{\mathcal E}_{n+1}^2}\left(1 - \frac{p_{\parallel}^2}{{\mathcal E}_{n+1}^2}\right) + f^S_n \frac{{\mathcal E}_n}{\omega^2 -4{\mathcal E}_n^2}\left( 1 - \frac{p_{\parallel}^2}{{\mathcal E}_n^2} \right) \right\}, \nonumber \\
\Pi^{00} & = &\Pi^{01} = \Pi^{02} = \Pi^{03} = \Pi^{13} = \Pi^{23} = 0,
\end{eqnarray}

\noindent where $f^S_n \equiv n^+({\mathcal E}_n) + n^-({\mathcal E}_n), f^D_n \equiv n^+({\mathcal E}_n) - n^-({\mathcal E}_n)$ and  ${\mathcal E}_n \equiv (m^2 + p_{\parallel}^2 +2neB)^{\frac{1}{2}}$, noting that $n^+$ is the electron occupation number and $n^-$ is the positron occupation number.  

Application of the long wavelength limit to $\Pi^{\mu\nu}_{\;\;5}$ yields

\begin{eqnarray}
\Pi^{11}_{\;\:5} &=& \Pi^{22}_{\;\:5} =\frac{e^3 B}{2\pi} \sum_{n=0}^{\infty} \int\frac{dp_{\parallel}}{2\pi} p_{\parallel}\omega\left(\frac{{\mathcal E}_{n+1} + {\mathcal E}_n}{{\mathcal E}_{n+1}{\mathcal E}_n}\right)(f^D_{n+1}-f^D_n), \nonumber \\
\Pi^{03}_{\;\:5} &=& \frac{2e^3 B}{2\pi} \sum_{n=0}^{\infty}\int\frac{dp_{\parallel}}{2\pi}\Bigg\{\left(1-\frac{(m^2+p_{\parallel}^2)}{{\mathcal E}_{n+1}^2}\right)f^S_{n+1}\frac{{\mathcal E}_{n+1}}{\omega^2-4{\mathcal E}_{n+1}^2} \vspace{0.2cm}  \nonumber \\ & & -\ \left(1-\frac{(m^2+p_{\parallel}^2)}{{\mathcal E}_n^2}\right)f^S_n\frac{{\mathcal E}_n}{\omega^2 - 4{\mathcal E}_n^2} \Bigg\}, \nonumber \\
\Pi^{30}_{\;\:5} &=& \frac{2e^3 B}{2\pi}\sum_{n=0}^{\infty}\int\frac{dp_{\parallel}}{2\pi}\Bigg\{ \left(1 +\frac{(m^2-p_{\parallel}^2)}{{\mathcal E}_{n+1}^2}\right)f^S_{n+1}\frac{{\mathcal E}_{n+1}}{\omega^2-{\mathcal E}_{n+1}^2} \vspace{0.1cm} \nonumber \\ 
& & -\ \left(1+\frac{(m^2 - p_{\parallel}^2)}{{\mathcal E}_n^2}\right) f^S_n\frac{{\mathcal E}_n}{\omega^2-4{\mathcal E}_n^2} \Bigg\}, \nonumber \\
\Pi^{00}_{\;\:5} &=& \Pi^{01}_{\;\:5} \  =  \ \Pi^{02}_{\;\:5} \ = \ \Pi^{12}_{\;\:5}\  = \ \Pi^{13}_{\;\:5} \ = \ \Pi^{23}_{\;\:5} \ = \ \Pi^{33}_{\;\:5} \ = \ 0 .
\end{eqnarray} 

\noindent By inspection, it is clear that if the electron and positron distribution functions are even functions of $p_{\parallel}$ then the $\Pi^{11}_{\;\:5}$ and $\Pi^{22}_{\;\:5}$ terms are identically zero.   Considering the other two components, $\Pi^{03}_{\;\:5}$ and $\Pi^{30}_{\:\;5}$, rearrangement of the sum over $n$ shows that only the $n = 0$ term can contribute.  The $n=0$ term vanishes identically for $\Pi^{03}_{\;\:5}$, which means that the only non-zero component of the axial vector response tensor is $\Pi^{30}_{\:\;5}$. 

\section{}
The results required to determine the response of a thermal magnetized plasma are summarized below.  The only $J$ functions that are required under the assumptions made in Section III are $J_0^0$ and $J_1^0$.  These have the simple forms

\begin{eqnarray}
J^0_0(u) & = & e^{-\frac{1}{2}u} , \\
J^0_1(u) & = & \sqrt{u}\ e^{-\frac{1}{2}u} ,
\end{eqnarray}
which are used below.  The ``non-resonant'' contributions to the response tensor are 

\begin{eqnarray}
\Pi^{00}_{\rm nr} & = & \varepsilon_0 \omega_p^2 \left[ \frac{k_{\parallel}^2}{4m^2} + \frac{k_{\perp}^2}{4m^2} \right]\exp\left(-\frac{p_{\parallel}^2}{2mT}\right), \nonumber 
\\
\Pi^{11}_{\rm nr} & = & \Pi^{22}_{\rm nr} = \varepsilon_0 \omega_p^2 \left[ 1 + \frac{\Omega_e}{2m} + \frac{k_{\parallel}^2}{4m^2}\right] \exp\left(-\frac{p_{\parallel}^2}{2mT}\right), \nonumber 
\\
\Pi^{33}_{\rm nr} & = & \varepsilon_0\omega_p^2\left[ 1 + \frac{\Omega_e}{2m} + \frac{k_{\parallel}^2}{4m^2} + \frac{k_{\perp}^2}{2eB}\right] \exp\left(-\frac{p_{\parallel}^2}{2mT}\right), \nonumber 
\\
\Pi^{01}_{\rm nr} & = & 0, \nonumber
\\
\Pi^{02}_{\rm nr} & = & i\varepsilon_0\omega_p^2\frac{k_{\perp}}{2m} \exp\left(-\frac{p_{\parallel}^2}{2mT}\right), \nonumber 
\\
\Pi^{03}_{\rm nr} & = &  0, \nonumber 
\\
\Pi^{12}_{\rm nr} & = & 0, \nonumber
\\
\Pi^{13}_{\rm nr} & = & -\varepsilon_0\omega_p^2 \frac{k_{\perp}k_{\parallel}}{4m^2} \exp\left(-\frac{p_{\parallel}^2}{2mT}\right), \nonumber 
\\
\Pi^{23}_{\rm nr} & = & i\varepsilon_0\omega_p^2 \frac{k_{\perp}k_{\parallel}}{4m^2} \exp\left(-\frac{p_{\parallel}^2}{2mT}\right),
\end{eqnarray}

\noindent and for the axial vector response,

\begin{eqnarray}
{\Pi^{00}_{\;\;5}}_{\rm nr} & = & {\Pi^{33}_{\;\;5}}_{\rm nr} = 0, \nonumber
\\
{\Pi^{11}_{\;\;5}}_{\rm nr} & = & {\Pi^{22}_{\;\;5}}_{\rm nr} = -\varepsilon_0\omega_p^2\frac{k_{\parallel}}{2m} \exp\left(-\frac{p_{\parallel}^2}{2mT}\right), \nonumber 
\\
{\Pi^{01}_{\;\;5}}_{\rm nr} & = & -\varepsilon_0\omega_p^2\frac{k_{\perp}k_{\parallel}}{4m^2} \exp\left(-\frac{p_{\parallel}^2}{2mT}\right), \nonumber 
\\
{\Pi^{02}_{\;\;5}}_{\rm nr} & = & -i\varepsilon_0\omega_p^2\frac{k_{\perp}k_{\parallel}}{4m^2} \exp\left(-\frac{p_{\parallel}^2}{2mT}\right), \nonumber 
\\
{\Pi^{03}_{\;\;5}}_{\rm nr} & = & -\varepsilon_0\omega_p^2\left[\frac{k_{\parallel}^2}{4m^2} + \frac{k_{\perp}^2}{4m^2}\right] \exp\left(-\frac{p_{\parallel}^2}{2mT}\right), \nonumber 
\\
{\Pi^{12}_{\;\;5}}_{\rm nr} & = & 0, \nonumber 
\\
{\Pi^{13}_{\;\;5}}_{\rm nr} & = & 0, \nonumber
\\
{\Pi^{23}_{\;\;5}}_{\rm nr} & = & i\varepsilon_0 \omega_p^2 \frac{k_{\perp}}{2m} \exp\left(-\frac{p_{\parallel}^2}{2mT}\right).
\end{eqnarray}

\noindent Due to the simplifications made in assuming a non-relativistic non-degenerate plasma, only one type of integral needs to be evaluated, namely

\begin{equation}
\label{def}
I_l = \int_{-\infty}^{\infty} dp_{\parallel} \frac{p_{\parallel}^l}{a_{\pm} - bp_{\parallel}}\exp\left(-\frac{p_{\parallel}^2}{2mT}\right), \quad\quad l = 0,1\, {\rm or} \, 2,
\end{equation}

\noindent where $T$ is the temperature, $a_{\pm}= \omega \pm \Omega_e$ and $b = k_{\parallel}/m$.  The integral in Eq.~(\ref{def}) may be re-expressed in terms of the plasma dispersion function $\bar{\phi}(z)$ defined in Eq.~(\ref{fizbuz}), which allows one to obtain the following forms for the $I_l$:

\begin{eqnarray}
 {\displaystyle I_0  =  \frac{\sqrt{2\pi mT}}{a_{\pm}}\bar{\phi}(z_{\pm}) }, \quad\quad
{\displaystyle I_1  =  \frac{\sqrt{2\pi mT}}{b} (\bar{\phi}(z_{\pm}) - 1)} , \\ \vspace{0.3cm} {\displaystyle I_2  =  \frac{a_{\pm}\sqrt{2\pi mT}}{b^2}(\bar{\phi}(z_{\pm}) -1)} ,
\end{eqnarray}
where $z_{\pm} = a_{\pm}/b\sqrt{2mT}$.  

Using Eq.~(\ref{spin}), the expressions in the resonant part of the response tensor may be evaluated 
\begin{eqnarray}
\Pi^{00}_{\rm res} & = & -\varepsilon_0 \omega_p^2   \frac{k_{\perp}^2}{ 2\Omega_e} \left[ \left(1 + \frac{\Omega_e}{2m} + \frac{k_{\parallel}^2}{4m^2}\right)\left(\frac{\bar{\phi}(z_+)}{\omega + \Omega_e} - \frac{\bar{\phi}(z_-)}{\omega - \Omega_e}\right) - \frac{1}{m}[\bar{\phi}(z_+) - \bar{\phi}(z_-)] \right]\exp\left(-\frac{k_{\perp}^2}{2eB}\right) , \nonumber 
\\
\Pi^{11}_{\rm res} & = & \Pi^{22}_{\rm res} = -\varepsilon_0 \omega_p^2 \left(\frac{\Omega_e}{2} + \frac{k_{\parallel}^2}{4m}\right)\left(\frac{\bar{\phi}(z_+)}{\omega + \Omega_e} - \frac{\bar{\phi}(z_-)}{\omega - \Omega_e}\right) \exp\left(-\frac{k_{\perp}^2}{2eB}\right) , \nonumber 
\\
\Pi^{33}_{\rm res} & = & -\varepsilon_0 \omega_p^2 \frac{k_{\perp}^2}{2\Omega_e}\left[\left(\frac{\Omega_e}{2m} + \frac{k_{\parallel}^2}{4m^2}\right)\left(\frac{\bar{\phi}(z_+)}{\omega + \Omega_e} - \frac{\bar{\phi}(z_-)}{\omega - \Omega_e}\right) - \frac{1}{m}[\bar{\phi}(z_+) - \bar{\phi}(z_-)] \right] \exp\left(-\frac{k_{\perp}^2}{2eB}\right) , \nonumber
\\
\Pi^{01}_{\rm res} & = & \varepsilon_0 \omega_p^2 \frac{k_{\perp}}{2}\left( \frac{\bar{\phi}(z_+)}{\omega + \Omega_e} + \frac{\bar{\phi}(z_-)}{\omega - \Omega_e} \right) \exp\left(-\frac{k_{\perp}^2}{2eB}\right) , \nonumber
\\
\Pi^{02}_{\rm res} & = & i\varepsilon_0 \omega_p^2 \frac{k_{\perp}}{2}\left( \frac{\bar{\phi}(z_+)}{\omega + \Omega_e} - \frac{\bar{\phi}(z_-)}{\omega - \Omega_e} \right) \exp\left(-\frac{k_{\perp}^2}{2eB}\right) , \nonumber 
\\
\Pi^{03}_{\rm res} & = & -\varepsilon_0 \omega_p^2 \frac{k_{\perp}^2}{2\Omega_e} \left[\frac{1}{k_{\parallel}}[\bar{\phi}(z_+) - \bar{\phi}(z_-)] - \frac{k_{\parallel}}{2m}\left(\frac{\bar{\phi}(z_+)}{\omega + \Omega_e} - \frac{\bar{\phi}(z_-)}{\omega - \Omega_e}\right)\right] \exp\left(-\frac{k_{\perp}^2}
{2eB}\right) , \nonumber
\\
\Pi^{12}_{\rm res} & = & -i\varepsilon_0\omega_p^2 \left(\frac{\Omega_e}{2} + \frac{k_{\parallel}^2}{4m^2}\right) \left( \frac{\bar{\phi}(z_+)}{\omega + \Omega_e} + \frac{\bar{\phi}(z_-)}{\omega - \Omega_e} \right)\exp\left(-\frac{k_{\perp}^2}{2eB}\right) , \nonumber
\\
\Pi^{13}_{\rm res} & = & \varepsilon_0 \omega_p^2 \frac{k_{\perp}}{2m} \left(\frac{m}{k_{\parallel}}[\bar{\phi}(z_+) + \bar{\phi}(z_-) - 2] - \frac{k_{\parallel}\bar{\phi}(z_-)}{\omega - \Omega_e}\right) \exp\left(-\frac{k_{\perp}^2}{2eB}\right) , \nonumber
\\
\Pi^{23}_{\rm res} & = & i\varepsilon_0 \omega_p^2 \frac{k_{\perp}}{2m} \left(\frac{m}{k_{\parallel}}[\bar{\phi}(z_+) - \bar{\phi}(z_-)] + \frac{k_{\parallel}\bar{\phi}(z_-)}{\omega - \Omega_e}\right) \exp\left(-\frac{k_{\perp}^2}{2eB}\right) 
 ,
\end{eqnarray}
and for the axial vector response

\begin{eqnarray}
{\Pi^{00}_{\;\;5}}_{\rm res} & = & {\Pi^{33}_{\;\;5}}_{\rm res} = -\varepsilon_0\omega_p^2\frac{k_{\perp}^2}{2\Omega_e}\left[\frac{1}{k_{\parallel}}\left( 1 + \frac{\Omega_e}{2m}\right)\left[\bar{\phi}(z_+) - \bar{\phi}(z_-)\right] - \frac{k_{\parallel}}{m}\left(\frac{\bar{\phi}(z_+)}{\omega + \Omega_e} - \frac{\bar{\phi}(z_-)}{\omega - \Omega_e}\right)\right]
\exp\left(-\frac{k_{\perp}^2}{2eB}\right) , \nonumber
\\
{\Pi^{11}_{\;\;5}}_{\rm res} & = & {\Pi^{22}_{\;\;5}}_{\rm res} = \varepsilon_0 \omega_p^2 \frac{1}{2} \left[ k_{\parallel} \left(\frac{\bar{\phi}(z_+)}{\omega + \Omega_e} - \frac{\bar{\phi}(z_-)}{\omega - \Omega_e}\right) + \frac{\Omega_e}{k_{\parallel}}[\bar{\phi}(z_+) - \bar{\phi}(z_-)]\right] \exp\left(-\frac{k_{\perp}^2}{2eB}\right) , \nonumber
\\
{\Pi^{01}_{\;\;5}}_{\rm res} & = & -\varepsilon_0\omega_p^2\frac{k_\perp}{2m}\left[\frac{m}{k_\parallel}[\bar{\phi}(z_+) + \bar{\phi}(z_-) - 2] - \frac{k_{\parallel}\bar{\phi}(z_-)}{\omega - \Omega_e}\right] \exp\left(-\frac{k_{\perp}^2}{2eB}\right) , \nonumber
\\
{\Pi^{02}_{\;\;5}}_{\rm res} & = & -i\varepsilon_0\omega_p^2 \frac{k_{\perp}}{2m}\left[\frac{m}{k_{\parallel}}[\bar{\phi}(z_+) - \bar{\phi}(z_-)] + \frac{k_{\parallel}\bar{\phi}(z_-)}{\omega - \Omega_e}\right]
\exp\left(-\frac{k_{\perp}^2}{2eB}\right) , \nonumber
\\
{\Pi^{03}_{\;\;5}}_{\rm res} & = & \varepsilon_0 \omega_p^2 \frac{k_{\perp}^2}{2\Omega_e}\left[\left(1 + \frac{\Omega_e}{2m} + \frac{k_{\parallel}^2}{4m^2}\right)\left(\frac{\bar{\phi}(z_+)}{\omega + \Omega_e} - \frac{\bar{\phi}(z_-)}{\omega - \Omega_e}\right)\right. \nonumber \\ & & \left.+ \frac{1}{k_{\parallel}^2} \left[ (\omega + \Omega_e)(\bar{\phi}(z_+) - 1) - (\omega - \Omega_e)(\bar{\phi}(z_-) - 1)\right]\right] \exp\left(-\frac{k_{\perp}^2}{2eB}\right) , \nonumber
\\
{\Pi^{12}_{\;\;5}}_{\rm res} & = & i\varepsilon_0 \omega_p^2 \frac{1}{2} \left[ k_{\parallel} \left(\frac{\bar{\phi}(z_+)}{\omega + \Omega_e} - \frac{\bar{\phi}(z_-)}{\omega - \Omega_e}\right) + \frac{\Omega_e}{k_{\parallel}}[\bar{\phi}(z_+) - \bar{\phi}(z_-)]\right] \exp\left(-\frac{k_{\perp}^2}{2eB}\right)
 , \nonumber
\\
{\Pi^{13}_{\;\;5}}_{\rm res} & = & -\varepsilon_0 \omega_p^2 \frac{k_{\perp}}{2}\left(\frac{\bar{\phi}(z_+)}{\omega + \Omega_e} + \frac{\bar{\phi}(z_-)}{\omega - \Omega_e}\right)
\exp\left(-\frac{k_{\perp}^2}{2eB}\right) , \nonumber
\\
{\Pi^{23}_{\;\;5}}_{\rm res} & = & i\varepsilon_0 \omega_p^2 \frac{k_{\perp}}{2}\left(\frac{\bar{\phi}(z_+)}{\omega + \Omega_e} - \frac{\bar{\phi}(z_-)}{\omega - \Omega_e}\right)
\exp\left(-\frac{k_{\perp}^2}{2eB}\right)
 , \nonumber
\\
{\Pi^{30}_{\;\;5}}_{\rm res} & = & \varepsilon_0 \omega_p^2\frac{k_{\perp}^2}{2\Omega_e}\left(\frac{\Omega_e}{2m} + \frac{k_{\parallel}^2}{4m^2}\right) \left(\frac{\bar{\phi}(z_+)}{\omega + \Omega_e} - \frac{\bar{\phi}(z_-)}{\omega - \Omega_e}\right)
\exp\left(-\frac{k_{\perp}^2}{2eB}\right).
\end{eqnarray}

Finally, we display the result for $\Pi^{\mu\nu}$ when one makes a high $z_{\pm}$ expansion, corresponding to $T \ll m$.  (The result for $\Pi^{\mu\nu}_5$ may be calculated similarly).

\begin{eqnarray}
\label{final}
\Pi^{00} & = & \varepsilon_0\omega_p^2 \left[ \left(\frac{k_{\parallel}^2}{4m^2} + \frac{k_{\perp}^2}{4m^2}\right) + \frac{k_{\perp}^2}{\omega^2 - \Omega_e^2}\left( 1 + \frac{T}{m} + \frac{\Omega_e}{2m} + \frac{k_{\parallel}^2}{4m^2}\right)\right] \exp\left(-\frac{k_{\perp}^2}{2eB}\right), \nonumber
\\
\Pi^{11} & = & \Pi^{22} = \varepsilon_0\omega_p^2\left[\frac{\omega^2}{\omega^2 - \Omega^2}
+ \frac{\Omega_e}{2m} + \frac{k_{\parallel}^2}{4m^2} +\frac{k_{\parallel}^2\Omega_e}{2m(\omega^2 - \Omega^2)}\right] 
\exp\left(-\frac{k_{\perp}^2}{2eB}\right), \nonumber
\\
\Pi^{33} & = & \varepsilon_0\omega_p^2\left[ 1 + \frac{\Omega_e}{2m} + \frac{k_{\parallel}^2}{4m^2} + \frac{k_{\perp}^2}{2eB} + \frac{k_{\perp}^2}{\omega^2 - \Omega_e^2}\left( \frac{T}{m} + \frac{\Omega_e}{2m} + \frac{k_{\parallel}^2}{4m^2}\right)\right] \exp\left(-\frac{k_{\perp}^2}{2eB}\right), \nonumber
\\
\Pi^{01} & = & \varepsilon_0\omega_p^2 \frac{k_{\perp}\omega}{\omega^2 - \Omega_e^2}\exp\left(-\frac{k_{\perp}^2}{2eB}\right), \nonumber
\\
\Pi^{02} & = & i\varepsilon_0\omega_p^2\left[ \frac{k_{\perp}}{2m} - \frac{k_{\perp}\Omega_e}{\omega^2 - \Omega_e^2}\right] \exp\left(-\frac{k_{\perp}^2}{2eB}\right), \nonumber
\\
\Pi^{03} & = & -\varepsilon_0\omega_p^2\frac{k_{\parallel}k_{\perp}^2}{m(\omega^2 - \Omega_e^2)} \left[ 1 - \frac{T\omega}{\omega^2 - \Omega_e^2}\right] \exp\left(-\frac{k_{\perp}^2}{2eB}\right), \nonumber
\\
\Pi^{12} & = & -i\varepsilon_0\omega_p^2 \frac{\omega\Omega_e}{\omega^2 - \Omega_e^2}\left[ 1 + \frac{k_{\parallel}^2}{2m\Omega_e}\right] \exp\left(-\frac{k_{\perp}^2}{2eB}\right), \nonumber
\\
\Pi^{13} & = & -\varepsilon_0\omega_p^2\frac{k_{\perp}k_{\parallel}}{m}\left[ \frac{1}{4m} + \frac{1}{2(\omega- \Omega_e)} - \frac{T(\omega^2 + \Omega_e^2)}{(\omega^2 - \Omega_e^2)^2}\right] \exp\left(-\frac{k_{\perp}^2}{2eB}\right), \nonumber
\\
\Pi^{23} & = & i\varepsilon_0\omega_p^2\frac{k_{\perp}k_{\parallel}}{m}\left[ \frac{1}{4m} + \frac{1}{2(\omega - \Omega_e)} - \frac{2T\omega\Omega_e}{(\omega^2 - \Omega_e^2)^2}\right] \exp\left(-\frac{k_{\perp}^2}{2eB}\right).
\end{eqnarray}

\end{appendix}

\newpage
\begin{figure} 
\caption{The W diagram, Z diagram and the V-A diagram for the plasma neutrino process}
\end{figure}

\begin{figure}
\caption{Comparison of emission from the ordinary mode at $\theta = \pi/2$, both with and without the axial vector coupling.  The power is in units of W\,m$^{-3}$ and the electron number density is in units of m$^{-3}$.  The temperature is $10^8 \, {\rm K}$ and the magnetic field is 0.1 B$_{\rm c}$.}
\end{figure}

\begin{figure}
\caption{Temperature and magnetic field regimes for which the axial vector coupling is likely to be important in astrophysical objects.}
\end{figure}

\newpage

\begin{table}
\caption{The natural modes of a cold plasma}
\vspace{0.1cm}
{\bf Modes at $\theta = 0$}, $\mbox{\boldmath $\kappa$} = (0,0,1)$ 
\begin{tabular}{lcc} 
Mode & Dispersion Relation & Polarization Vector \\ \hline 
\, & \, & \, \\
Longitudinal & $\displaystyle{P=0}$ & ${\mathbf e} = (0,0,1)$ \\
\, &\, &\, \\
Ordinary & $n_M^2 = S + D$ & ${\mathbf e} = {\displaystyle \frac{1}{\sqrt{2}}(1,i,0)}$ \\
\, & \, & \, \\
Extraordinary & $n_M^2 = S - D$ & ${\mathbf e} = {\displaystyle \frac{1}{\sqrt{2}}(1,-i,0)}$ \\
\end{tabular} 
\vspace{0.3cm}
{\bf Modes at $\theta = \pi/2$}, $\mbox{\boldmath $\kappa$} = (1,0,0)$ 
\begin{tabular}{lcc}  
Mode & Dispersion Relation & Polarization Vector \\ \hline 
\, & \, & \, \\
Ordinary & $\displaystyle{n_M^2 = P}$ & ${\mathbf e} = (0,0,1)$ \\
\, & \, & \, \\
Extraordinary & $n_M^2 = {\displaystyle \frac{S^2-D^2}{S}}$ & ${\mathbf e} = {\displaystyle \frac{(D,-iS,0)}{\sqrt{D^2 + S^2}}}$ \\
\end{tabular}
\end{table}

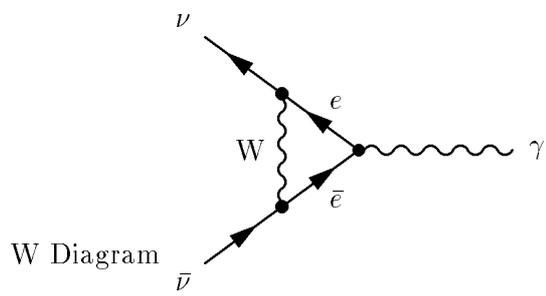
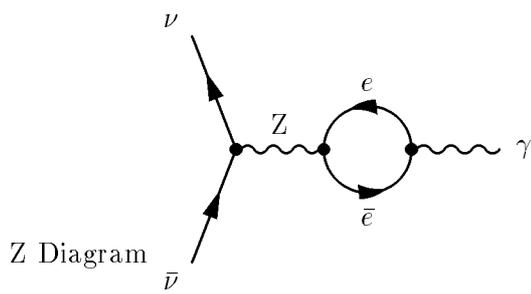
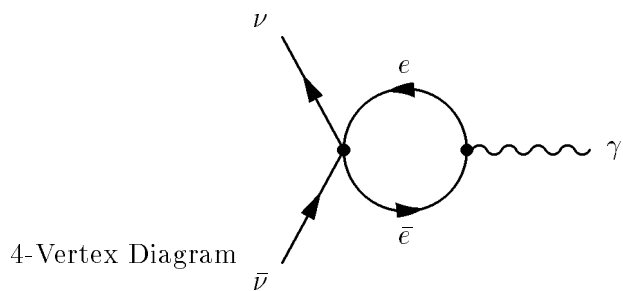
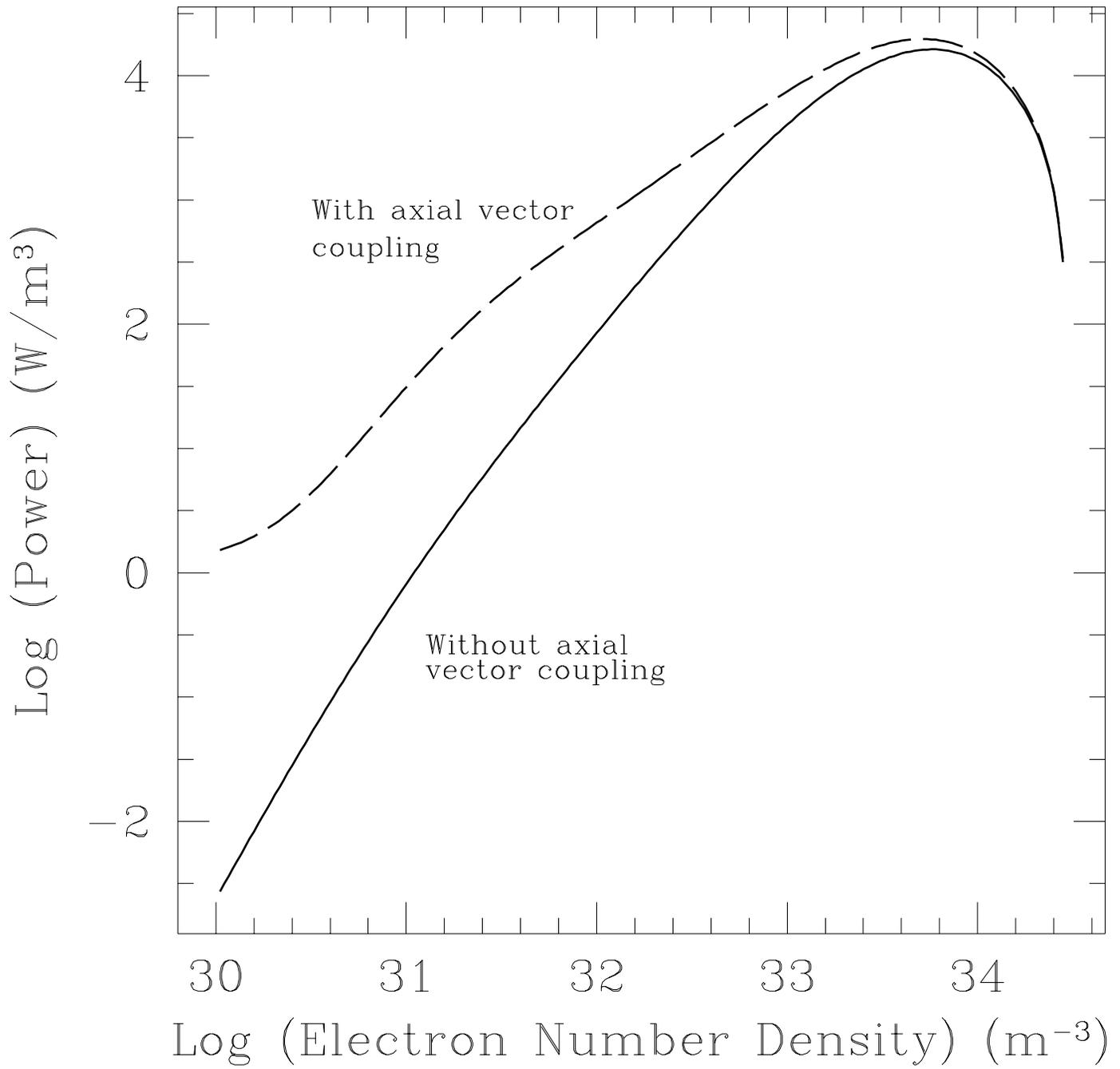
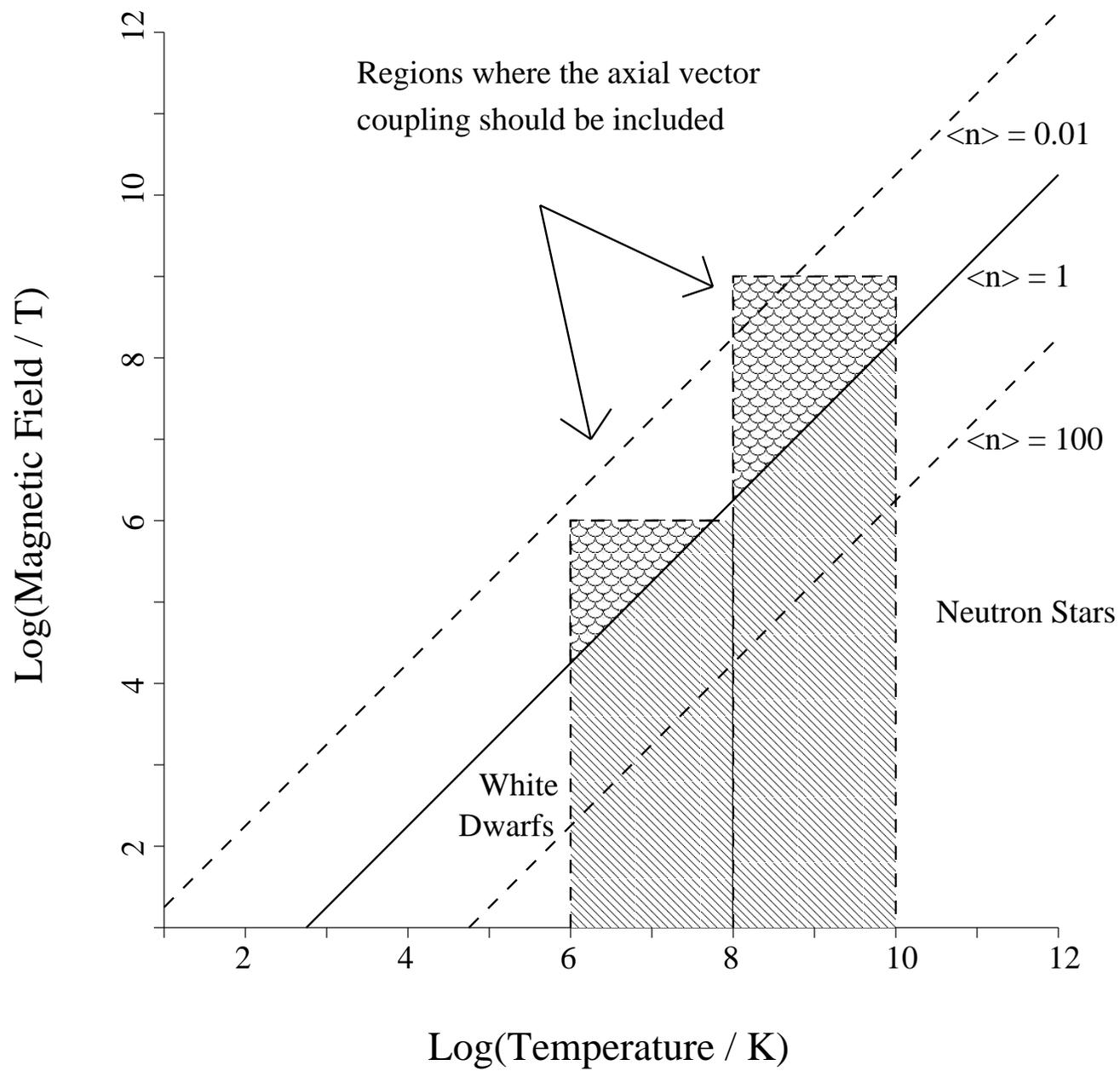
\end{document}